\documentclass[10pt,conference]{IEEEtran}
\IEEEoverridecommandlockouts
\usepackage{xspace}
\usepackage{caption}
\usepackage{pdfpages}
\usepackage{hyperref}
\usepackage{amsmath,amsfonts}
\usepackage{algorithmic}
\usepackage{graphicx}
\usepackage[utf8]{inputenc} 
\usepackage{textcomp}
\usepackage{listings}
\usepackage{xcolor, soul}
\usepackage[all]{nowidow}
\usepackage[inline]{enumitem}
\usepackage[scaled]{beramono}
\usepackage{multirow}
\usepackage{tikz,pifont}
\usepackage{graphicx} 
\usepackage{grffile}
\usepackage{pgfplots}
\usepackage{cleveref}
\usepackage{color, colortbl}
\usepackage[many]{tcolorbox}
\usepackage{pdflscape}
\usepackage{makecell}
\usepackage{pbox}
\usetikzlibrary{shadows}
\usepackage{balance}
\usepackage{ragged2e}
\usepackage[tight,footnotesize]{subfigure}
\usepackage{amssymb}
\usepackage{fontawesome}

\pgfplotsset{compat=1.8}
\usepgfplotslibrary{statistics}

\colorlet{lightgray}{gray!30}

\sethlcolor{lightgray}

\definecolor{mygreen}{rgb}{0,0.6,0}
\definecolor{mygray}{rgb}{0.95,0.95,0.95}
\definecolor{myred}{rgb}{0.5,0,0}

\definecolor{Gray}{gray}{0.9}

\newcommand\untick{\ding{54}}

\usepackage{amssymb}

\newtcolorbox{shadedbox}{
	drop shadow southeast,
	breakable,
	enhanced jigsaw,
	colback=white,
	boxrule=0.80pt,
	left=0.3em,
	right=0.3em,
	top=0.1em,
	bottom=0.05em
}

\newcommand*{\ie}{i.e.,\@\xspace}
\newcommand*{\eg}{e.g.,\@\xspace}

\newcommand*{\GH}{GitHub\@\xspace}

\newcommand*{\CR}{CrossRec\@\xspace}
\newcommand*{\LR}{LibRec\@\xspace}
\newcommand*{\LS}{LibSeek\@\xspace}
\newcommand*{\GR}{GRec\@\xspace}

\newcommand*{\numPapers}{9,435\@\xspace}
\newcommand*{\numSys}{four\@\xspace}

\newcommand{\library}[1]{\texttt{\hl{\small #1}}}
\newcommand{\librarysmall}[1]{\texttt{\hl{\scriptsize #1}}}

\makeatletter
\newcommand*{\etc}{%
	\@ifnextchar{.}%
	{etc}%
	{etc.\@\xspace}%
}
\makeatother
\newcommand*{\etal}{\emph{et~al.}\@\xspace}
\newcommand\revised[1]{\textcolor{black}{#1}}

\newcommand{\rqfirst}{\textbf{RQ$_1$}: \emph{Has the issue of dealing with popularity bias in TPL recommender systems been studied by the existing literature?}} 
\newcommand{\rqsecond}{\textbf{RQ$_2$}: \emph{How well do state-of-the-art TPL recommender systems tackle popularity bias?}} 
\newcommand{\rqthird}{\textbf{RQ$_3$}: \emph{Can the re-ranking mechanism counteract popularity bias in TPL recommender systems?}}

\definecolor{verylightgray}{gray}{0.99}
\definecolor{lightgray}{gray}{0.92}

\definecolor{mygreen}{rgb}{0,0.6,0}
\definecolor{mygray}{rgb}{0.95,0.95,0.95}
\definecolor{myred}{rgb}{0.5,0,0}

\lstdefinestyle{JavaStyle} {
	backgroundcolor=\color{verylightgray},   
	commentstyle=\color{mygreen}, 
	breakatwhitespace=false,
	keywordstyle=\color{violet},
	language=Java,
	stringstyle=\color{blue},
	basicstyle=\scriptsize,
	showstringspaces=false
}

\lstdefinestyle{searchstringstyle}{
	basicstyle=\ttfamily\scriptsize,
	captionpos=t,                    
	numbers=none,                    
	numbersep=5pt,                  
	showspaces=false,                
	showstringspaces=false,
	showtabs=false,                  
	tabsize=2,
	frame=single
}

\begin{document}


\title{Dealing with Popularity Bias in \\Recommender Systems for Third-party Libraries: \\How far Are We?}






\author{\IEEEauthorblockN{Phuong T. Nguyen}
\IEEEauthorblockA{\textit{Universit\`a degli studi dell'Aquila, Italy} \\
phuong.nguyen@univaq.it} 
\and
\IEEEauthorblockN{Riccardo Rubei}
\IEEEauthorblockA{\textit{Universit\`a degli studi dell'Aquila, Italy} \\
	riccardo.rubei@graduate.univaq.it}
\and
\IEEEauthorblockN{Juri Di Rocco}
\IEEEauthorblockA{\textit{Universit\`a degli studi dell'Aquila, Italy} \\
	juri.dirocco@univaq.it}
\and
\IEEEauthorblockN{Claudio Di Sipio}
\IEEEauthorblockA{\textit{Universit\`a degli studi dell'Aquila, Italy} \\
	claudio.disipio@graduate.univaq.it}
\and
\IEEEauthorblockN{Davide Di Ruscio}
\IEEEauthorblockA{\textit{Universit\`a degli studi dell'Aquila, Italy} \\
	davide.diruscio@univaq.it}
\and
\IEEEauthorblockN{Massimiliano Di Penta} 
\IEEEauthorblockA{\textit{Universit\`a degli studi del Sannio, Italy} \\
dipenta@unisannio.it}
}


\maketitle

\begin{abstract}
Recommender systems for software engineering (RSSEs) assist software engineers in dealing with a growing information overload when discerning alternative development solutions. 
While RSSEs are becoming more and more effective in suggesting handy recommendations, they tend to suffer from popularity bias, \ie favoring items that are relevant mainly because several developers are using them. While this rewards artifacts that  are likely more reliable and well-documented, it would also mean that missing artifacts are rarely used because they are very specific or more recent. 
This paper studies popularity bias in Third-Party Library (TPL) RSSEs. First, we investigate whether state-of-the-art research in RSSEs has already tackled the issue of popularity bias. Then, we quantitatively assess \numSys existing TPL RSSEs, exploring their capability to deal with the recommendation of popular items. Finally, we propose a mechanism to defuse popularity bias in the recommendation list. The empirical study reveals that the issue of dealing with popularity in TPL RSSEs has not received adequate attention from the software engineering community. Among the surveyed work, only one starts investigating the issue, albeit getting a low prediction performance.

\end{abstract}


\thispagestyle{empty}

\section{Introduction}
\label{sec:Introduction}



In online shopping market, recommender systems have been conceived to enable customers to approach 
goods 
that suite their needs~\cite{DBLP:books/sp/Aggarwal16,DBLP:reference/rsh/2011}. 
Nonetheless, while being able to provide accurate results, these systems tend to present frequently seen items~\cite{DBLP:conf/flairs/AbdollahpouriBM19,DBLP:conf/recsys/AbdollahpouriMB19,10.1145/3564284}, 
\ie \emph{popularity bias} is a common phenomenon of various recommender systems. 
In fact, the ability to exhibit rare but useful items 
is considered as an added value~\cite{10.1145/1454008.1454012}. 
The \emph{long tail effect} indicates that a handful of items are extremely popular, whilst most of the remaining ones, so-called the long tail, are not known by customers \cite{Anderson:2006:LTW:1197299,4688070}. Essentially, products belonging to the long tail are considered a social good~\cite{DBLP:conf/recsys/AbdollahpouriMB19} and recommending them benefits both customers and shop owners \cite{Vargas_sales_diversity_14}.

Recommender systems for software engineering (RSSEs) effectively help developers cope with the proliferation of various data sources, and avoid information overload~\cite{di_rocco_development_2021,robillard_recommendation_2014}. RSSEs could suggest, for example, relevant developers' discussions, experts to handle issues or to fix bugs, code completions, or bug/vulnerability fixing. Among different types of RSSEs, in this paper, we focus on third-party library (TPL) RSSEs, that provide libraries relevant to the project under development~\cite{10.1007/978-3-030-64694-3_13,9043686,7985674,NGUYEN2019110460,SAIED2018164,LibRec}. Previous work has shown such systems to be highly accurate and valuable for developers. Furthermore, TPL RSSEs enable developers to use existing infrastructures to build software without reinventing the wheel by offering tailored, ready-to-be-used pieces of functionality.

In TPL recommendation, the \emph{Novelty} metric assesses if a system can retrieve libraries in the long tail and expose them to projects~\cite{NGUYEN2019110460}. This increases the possibility of coming across \emph{serendipitous} libraries \cite{Ge:2010_catalog_coverage}, \eg those that are seen by chance but turn out to be useful for the project under development~\cite{di_rocco_development_2021}. For example, there could be a recent library, yet to be widely used, that can better interface with new hardware or achieve faster performance than popular ones. In summary, recommending only popular TPLs would harm the novelty of the results. We conjecture that the existing research in RSSEs, while achieving enhancement in accuracy, neglects the issue of dealing with popularity bias in RSSEs.


In this paper, we perform a threefold investigation on the issue of popularity bias in TPL RSSEs, contributing to advance the knowledge in the domain as follows: 

\begin{itemize}
	\item[--] By means of a literature review on premier Software Engineering venues, we show that state-of-the-art research overlooks the issue of popularity bias in TPL RSSEs.  
	\item[--] We perform a 	qualitative analysis on recently developed RSSEs, studying their ability to deal with popularity bias.
	\item[--] Through a quantitative study on \numSys TPL RSSEs, we assess their sensitivity to popularity bias.
	\item[--] We propose a practical solution to increase fairness in the recommendation lists.  
\end{itemize}

The scripts and data produced in our study are available to foster future research~\cite{BiasInRSSErepo}.

\textbf{Structure.} Section \ref{sec:Background} presents a motivating example and background related to popularity bias. Section \ref{sec:Evaluation} elaborates on the study materials and methods. Afterward, the study outcomes are reported and analyzed in Section~\ref{sec:Results}. In Section~\ref{sec:Countermeasures}, we present empirical evidence on a possible counteraction. The related work is reviewed in Section~\ref{sec:RelatedWork}. We sketch future work and conclude our paper in Section~\ref{sec:Conclusion}. 

\section{Motivating Analysis} 
\label{sec:Background}




We motivate our work by analyzing the nature of the recommendations provided by a state-of-the-art TPL RSSE, \ie  \CR~\cite{NGUYEN2019110460}\footnote{The original implementation of \CR has been used.}  The system has been chosen for this motivating study because of: \emph{(i)}  original implementation availability and full documentation \cite{NGUYEN2019110460}, allowing us to monitor the internal 
process; \emph{(ii)} being less computationally-expensive than other tools such as  \LS~\cite{9043686} and \GR~\cite{10.1145/3468264.3468552}.

We are interested in understanding how \CR and similar RSSEs treat libraries having different frequencies of occurrences. To this end, 
we leveraged a dataset of generic Java applications with 5,200 projects using 31,817 libraries. To assess how different TPLs are recommended, we employed a typical recommendation scenario for each project: half of its used libraries are used as query, and the other half are removed to be the ground-truth data. Given the TPLs as query, \CR is expected to return the ones in the ground truth.

The ability to provide matching libraries is deemed to be among the most important quality traits of RSSEs~\cite{robillard_recommendation_2014}, including TPL RSSEs. Even though in practice a recommended library that does not belong to the ground-truth data might still be suitable for the project under consideration~\cite{Ge:2010_catalog_coverage,NGUYEN2019110460}. 
However, the usefulness, as well as the relevance of such a library needs to be carefully evaluated with a proper user study~\cite{di_rocco_development_2021}, and this is out of scope for this work. 

For each library, we consider the following parameters: $\mathbb{F}$ is the frequency of occurrences of a library in the training data; $\mathbb{G}$ is the number of projects containing the library in the ground-truth data; and  $\mathbb{R}$ is the number of projects getting the library as a recommendation by means of \CR. 


\begin{table}[h!]
	\centering
	\scriptsize	
	\caption{Libraries and \CR recommendations.}
	\begin{tabular}{|p{0.15cm}|p{3.5cm} | p{0.65cm} | p{0.65cm} | p{0.77cm} |}
		\hline
		                                                                     & \textbf{Library}                                     & $\mathbb{F}$ & $\mathbb{G}$ &$\mathbb{R}$ \\ \hline
		{\multirow{10}{*}{\rotatebox[origin=c]{90}{Popular}}}                & \texttt{\hl{\scriptsize junit:junit}}                & 3,926        & 3,468        & 4,615                     \\ \cline{2-5}
		                                                                     & \texttt{\hl{\scriptsize org.slf4j:slf4j-api}}        & 1,565        & 341          & 2,328                     \\ \cline{2-5}
		                                                                     & \texttt{\hl{\scriptsize log4j:log4j}}                & 1,106        & 890          & 2,592                     \\ \cline{2-5}
		                                                                     & \texttt{\hl{\scriptsize commons-io:commons-io}}      & 1,000        & 209          & 2,035                     \\ \cline{2-5}
		                                                                     & \texttt{\hl{\scriptsize com.google.guava:guava}}     & 956          & 414          & 2,105                     \\ \cline{2-5}
		                                                                     & \texttt{\hl{\scriptsize com.google.code.gson:gson}}  & 820          & 487          & 2,007                     \\ \cline{2-5}
		                                                                     & \texttt{\hl{\scriptsize org.mockito:mockito-core}}   & 689          & 548          & 1,658                     \\ \cline{2-5}
		                                                                     & \texttt{\hl{\scriptsize mysql:mysql-connector-java}} & 627          & 97           & 1,065                     \\ \cline{2-5}
		                                                                     & \texttt{\hl{\scriptsize joda-time:joda-time}}        & 513          & 314          & 1,054                     \\ \cline{2-5}
		                                                                     & \texttt{\hl{\scriptsize com.h2database:h2}}          & 410          & 325          & 588                       \\ \hline
		
		{\multirow{10}{*}{\rotatebox[origin=c]{90}{Rare}}} 
		                                                                     & \texttt{\hl{\scriptsize bsf:bsf}}                    & 12           & 9            & 0                         \\ \cline{2-5}
		                                                                     & \texttt{\hl{\scriptsize net.oauth.core:oauth}}       & 9            & 5            & 1                         \\ \cline{2-5}
		                                                                     & \texttt{\hl{\scriptsize simple-jndi:simple-jndi}}    & 8            & 7            & 0                         \\ \cline{2-5}
		                                                                     & \texttt{\hl{\scriptsize javax.jdo:jdo-api}}          & 7            & 4            & 0                         \\ \cline{2-5}
		                                                                     & \texttt{\hl{\scriptsize javax.el:javax.el}}          & 6            & 4            & 0                         \\ \cline{2-5}
		                                                                     & \texttt{\hl{\scriptsize org.joda:joda-money}}        & 6            & 4            & 0                         \\ \cline{2-5}
		                                                                     & \texttt{\hl{\scriptsize taglibs:request}}            & 5            & 1            & 0                         \\ \cline{2-5}
		                                                                     & \texttt{\hl{\scriptsize javatar:javatar}}            & 4            & 1            & 0                         \\ \cline{2-5}
		                                                                     & \texttt{\hl{\scriptsize batik:batik-codec}}          & 3            & 1            & 0                         \\ \cline{2-5}
		                                                                     & \texttt{\hl{\scriptsize org.neo4j:parent}}           & 2            & 1            & 0                         \\ \hline
	\end{tabular}
	\label{tab:MotivatingExample}
\end{table}



\subsection{Popular and rare libraries}
Table~\ref{tab:MotivatingExample} reports how recommendation results vary for the  10 most popular (\ie large  $\mathbb{F}$) and 10 least popular (\ie low $\mathbb{F}$) libraries in the dataset.
Table~\ref{tab:MotivatingExample} shows  that the most popular libraries are also recommended very often ($\mathbb{R}$ is large). Unsurprisingly, \texttt{\hl{\small junit:junit}} is the most popular library in the corpus as it appears $\mathbb{F}$=3,926 times in the training data.
Looking into the ground-truth data, we see that the library belongs to $\mathbb{G}$=3,468 projects. However, it is recommended to $\mathbb{R}$=4,615 projects, meaning that its popularity may generate several false positives. 
This is not particularly surprising, considering that one could think about filtering out from recommendations libraries not used in production code, as is the case of \texttt{\hl{\small junit:junit}}.
However, results are also true for other popular libraries, \ie they are frequently recommended by the RSSE, although they are not necessarily useful for several projects. Among others, 
 \texttt{\hl{\small mysql:mysql-connector-java}} belongs to the ground-truth data of $\mathbb{G}$=97 projects, however it is recommended $\mathbb{R}$=1,065 times. This 
 means that, most of the times, the recommendation turns out to be a false positive. We computed the Spearman's correlation coefficient and obtained 
 $\rho(\mathbb{F},\mathbb{R})$=0.976, \ie 
 there is a strong correlation between the frequency of a library usage and the number of times it gets recommended. Interestingly, 
 $\rho(\mathbb{G},\mathbb{R})$=0.29, and this corresponds to 
 a weak, and even no correlation between the number of times that a library appears in the ground-truth data and its frequency of recommendations. In other words, \emph{a popular library is recommended just because it is popular, not because it is needed by projects}.


The bottom of Table~\ref{tab:MotivatingExample} highlights a striking outcome: almost all the rare TPLs are never discovered by \CR, \ie $\mathbb{R}$=0. For example, \texttt{\hl{\small simple-jndi:simple-jndi}} belongs to eight and seven projects in the training and ground-truth data, respectively, however, it is never recommended. This means that projects such as \texttt{\hl{\small apache/openjpa}} that contain \texttt{\hl{\small simple-jndi:simple-jndi}} in their ground-truth data will never get such a library recommended. 
In summary, the rare TPLs in Table~\ref{tab:MotivatingExample} get almost no chance to be recommended.


\vspace{.1cm}
\begin{shadedbox}
	\small{$\star$ \textbf{Remark 1.} Popular libraries are excessively recommended to projects, leading to false positives.  Given that TPL RSSEs present only popular libraries, the rare ones will never be recommended.}
\end{shadedbox}

\begin{table*}[t!]
	\centering
	\scriptsize	
	\caption{Explanatory \CR True Positive and False Negative recommendations.}
	\begin{tabular}{|p{2.1cm}|p{7.3cm} |p{7.3cm} |}	\hline
		\textbf{Project}& \textbf{True Positive} & \textbf{False Negative} \\ \hline
		\librarysmall{org.eclipse.kapua} & \librarysmall{commons-fileupload:commons-fileupload}, \librarysmall{com.h2database:h2}, \librarysmall{org.elasticsearch.client: transport}, \librarysmall{commons-io:commons-io}, ... & \librarysmall{org.apache.tomcat: tomcat-servlet-api}, \librarysmall{org.reflections:reflections}, \librarysmall{org.apache.activemq: artemis-amqp-protocol} ... 	\\	\hline
		\librarysmall{org.eclipse.mylyn} & \librarysmall{junit:junit} \librarysmall{com.google.code.gson: gson}, ... & \librarysmall{org.eclipse.emf.ecore.xmi}, \librarysmall{org.eclipse.emf.common}, \librarysmall{org.eclipse.emf.ecore}, \librarysmall{org.eclipse.core.runtime}, ... \\ \hline
		\librarysmall{hadoop.apache.org} &  \librarysmall{org.hsqldb:hsqldb}, \librarysmall{org.mockito:mockito-all}, \librarysmall{commons-daemon:commons-daemon},  \librarysmall{commons-net:commons-net}, ... & \librarysmall{woodstox-core}, \librarysmall{kerb-simplekdc}, ...\\ \hline		
	\end{tabular}
	\label{tab:MotivatingExample-2}
\end{table*}

\subsection{Solution-specific libraries} 
As previously mentioned, RSSEs suggest TPLs by relying on some definitions of project similarity. Unfortunately, this might not be sufficient to recommend TPLs specific to peculiar goals or used technology of a project at hand. Table \ref{tab:MotivatingExample-2} shows examples of true positive and false negative recommendations produced by \CR for some explanatory projects.

Eclipse Kapua \cite{kapua} is an integration platform for IoT devices and smart sensors. Thus, it manages the connectivity of different devices in heterogeneous environments. By asking \CR to recommend libraries for \library{org.eclipse.kapua} by intentionally removing some actual dependencies, we only get generic libraries, like \library{commons-io} and \library{spring-security-core}. Unfortunately, libraries specific to the project's goal, \ie managing connectivity and service development, such as \library{tomcat-servlet-api} and \library{artemis-amqp-protocol}, are not recommended. 

Eclipse Mylyn \cite{mylyn} is the task and application lifecycle management (ALM) framework for Eclipse. \CR can recommend for such a project generic libraries like \library{junit:junit}, and \library{com.google.code.gson:gson}, whereas it fails to recommend libraries that are related to the Eclipse architecture such as OSGi \cite{osgi} and EMF \cite{emf} specific libraries.

Apache Hadoop \cite{hadoop} is a framework for managing the distributed processing of large data sets across clusters of computers. Also in this case, \CR can correctly recommend libraries such as \library{commons-daemon}, \library{common-net}. In contrast, it is not able to provide 
useful libraries 
related to the specific goal of the project, like fast processing of XML documents (\library{woodstox-core}) and distributed authentication (\library{kerb-simplekdc}).

\begin{shadedbox}
	\small{$\star$ \textbf{Remark 2.} As found for \CR, state-of-the-art TPL RSSEs may not properly leverage peculiar aspects of a project, \eg related to implementation solutions. Hence, they fail to recommend some relevant goals, architecture- or solution-specific libraries.}
\end{shadedbox}

Altogether, we conclude that \emph{popularity bias does exist in TPL RSSEs}. 
Such an issue needs to be tackled in order to decrease the number of frequent (while not necessarily useful) libraries, and to increase the number of the rare (but useful) ones in the recommendations. 


\section{Investigation materials and methods}
\label{sec:Evaluation}

The \emph{goal} of this study is to investigate the extent to which TPL RSSEs have addressed popularity bias. The \emph{quality focus} concerns the accuracy of the provided recommendations, which can be affected by the presence of popular artifacts in the training data. Finally, the \emph{context} consists of \emph{(i)} TPL RSSEs published in the recent software engineering literature; and \emph{(ii)} three datasets from Java and Android open-source projects used to assess the recommenders.





This study addresses the following research questions:


\begin{itemize}
	\item \rqfirst~Through a literature analysis, we investigate to what extent the popularity bias in TPL RSSEs has ever been studied by state-of-the-art research. Though we do not aim for a complete, detailed systematic literature review, we adhere to existing guidelines for such type of study in Software Engineering research \cite{KitchenhamBLBB11,MacDonellSKM10,DBLP:conf/ease/Wohlin14}.
	\item \rqsecond~We analyze the most representative TPL RSSEs to investigate the extent to which they are susceptible to popularity bias. Afterward, we perform an empirical evaluation on four representative systems, \ie \LR~\cite{LibRec}, \CR~\cite{NGUYEN2019110460}, \LS~\cite{9043686}, and \GR~\cite{10.1145/3468264.3468552} further validating our hypothesis. Their source implementation is available, 
	allowing us to experiment on real-world datasets according to our needs. 
	\item \rqthird~Based on an existing Web search diversification technique~\cite{10.1145/1772690.1772780}, we derive a re-ranking strategy to defuse bias in the recommendation results provided by two TPL RSSEs. 
\end{itemize}



\subsection{RQ$_1$ Methodology: Literature analysis} \label{sec:Collection}

	\begin{lstlisting}[label=lst:searchString,caption=Excerpt of the Scopus query string.,style=searchstringstyle]
		TITLE-ABS((``recommendation'' OR ``recommender'' OR ``recommendation systems'' OR ``recommender systems'')  
		AND (``library'' OR ``libraries'' OR ``TPL'' OR ``TPLs'' OR ``third-party libraries'' OR ``third-party'') [...] 
		AND (LIMIT-TO (PUBYEAR, 2022) OR [...]  OR  LIMIT-TO (PUBYEAR, 2017))  
	\end{lstlisting}

To address RQ$_1$, we perform a literature analysis to gain a first understanding, based on the description available in scientific papers, of the extent to which the bias problem has been addressed for RSSEs. Aiming at a reasonable trade-off between efficiency and the coverage of state-of-the-art studies on popularity bias in TPL RSSEs, we employed a search strategy guided by four W-questions~\cite{zhang2011identifying}, \ie ``\emph{Which?}'' ``\emph{Where?}'' ``\emph{What?}'' ``\emph{When?}'' explained as follows. 

\begin{itemize}
	\item \emph{Which?} We conducted a combined search with automatic and manual activities to collect papers from various venues comprising conferences and journals. 
	\item \emph{Where?} The literature analysis was performed on premier software engineering venues including \emph{(i)} nine conferences: ASE, ESEC/FSE, ESEM, ICSE, ICSME, ICST, ISSTA, MSR, and SANER; and \emph{(ii)} five journals: EMSE, IST, JSS, TOSEM, and TSE.\footnote{Venues listed in alphabetical order, for full names, interested readers can refer to the online appendix \cite{BiasInRSSErepo}.} 
	The \textit{Scopus} database\footnote{\url{https://scopus.com}}  was chosen for the automatic search, and all the papers published by a given year of a given venue 
	were retrieved through the advanced Scopus search and export features. An excerpt of the queries is shown in Listing~\ref{lst:searchString}.
	\item \emph{What?} Title and abstract of each article were fetched following a set of predefined keywords. 
	\item \emph{When?} Since bias and fairness is a recent research theme, the search was confined to the most five recent years, \ie from 2017 to 2022. 
\end{itemize}

\vspace{-.1cm}
The search resulted in a corpus of \numPapers articles. Then, various filtering steps were conducted to scale down the search and retrieve only those that satisfy our predefined requirements. In particular, we are interested in studies dealing with RSSEs together with the relevant topics, \ie TPL recommendation, bias, and fairness. The metadata of the collected studies is available in an online appendix \cite{BiasInRSSErepo}. 

\subsection{RQ$_2$ Methodology: Quantitative analysis of popularity bias in \numSys TPL RSSEs} \label{sec:RQ2Methodology}

To address RQ$_2$, 
we select \numSys TPL RSSEs,  \ie \LR~\cite{LibRec}, \CR~\cite{NGUYEN2019110460}, \LS~\cite{9043686}, and \GR~\cite{10.1145/3468264.3468552}.
The rationale behind this selection is as follows. \LR and \CR are notable systems for recommending TPL libraries for generic applications, while \LS and \GR are devised to work with Android apps. The \numSys systems are also representative in terms of the working mechanisms. In particular, \LR\footnote{The original implementation of \LR was kindly provided by its authors.} and \CR\footnote{\url{https://github.com/crossminer/CrossRec}} are based on collaborative filtering techniques and associate rule mining; whilst \LS\footnote{\url{https://github.com/libseek/LibSeek}} employs matrix factorization, and \GR\footnote{\url{https://github.com/fio1982/GRec}} works on top of graph neural networks. We conjecture that, pending some possible external validity threats, an analysis of these systems could be generalizable to other TPL RSSEs.

\subsubsection{Datasets and configurations} \label{sec:Datasets}

We use datasets curated by existing studies to evaluate the \numSys 
TPL RSSEs. The datasets are summarized in Table~\ref{tab:DatasetsAndSystems}. Overall, we consider three datasets from Java open source software, two (DS$_1$ and DS$_2$) related to generic software systems, and one (DS$_3$) specific to Android apps.

\begin{table}[h!]
	\centering
	\scriptsize	
	\caption{Datasets and configurations.}
	\begin{tabular}{|p{1.40cm} | p{2.00cm} | p{2.00cm} | p{1.60cm} |}	\hline
		& \multicolumn{3}{c|}{\textbf{Dataset}} \\ \cline{2-4}
		& \textbf{DS$_1$}~\cite{NGUYEN2019110460}  & \textbf{DS$_2$}  & \textbf{DS$_3$}~\cite{9043686,10.1145/3468264.3468552}  \\ \hline 
		\# projects & 1,200 & 5,200 & 56,091 \\ 
		\# libraries & 13,497 & 31,817 & 762 \\ 
		Type & Generic software & Generic software & Android apps \\ \hline 
		\textbf{System} & \multicolumn{3}{c|}{\textbf{Experimental configurations}} \\ \hline
		\LR & \cellcolor{lightgray}\faCheck & \cellcolor{lightgray}\faCheck & \untick \\ 
		\CR & \cellcolor{lightgray}\faCheck & \cellcolor{lightgray}\faCheck & \untick \\ 
		\LS & \untick & \cellcolor{lightgray}\faCheck & \cellcolor{lightgray}\faCheck \\ 
		\GR & \untick & \cellcolor{lightgray}\faCheck & \cellcolor{lightgray}\faCheck \\ \hline		
	\end{tabular}
	\label{tab:DatasetsAndSystems}
\end{table}


\begin{figure*}[t!]
	\centering    
	\begin{tabular}{c c c}		
		\subfigure[\textbf{Dataset DS$_{1}$}]{\label{fig:D1}\includegraphics[width=55mm]{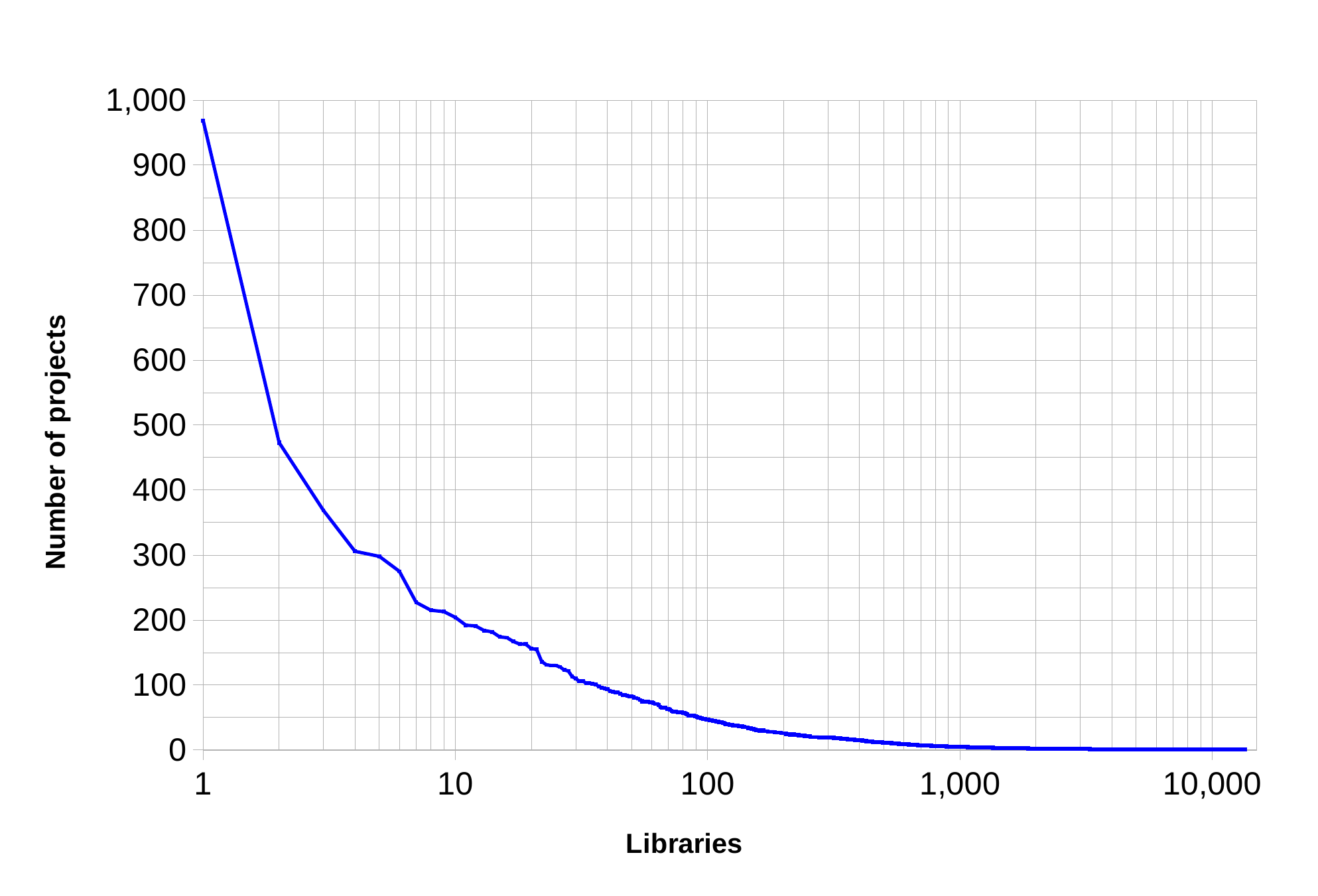}} &		
		\subfigure[\textbf{Dataset DS$_{2}$}]{\label{fig:D2}\includegraphics[width=55mm]{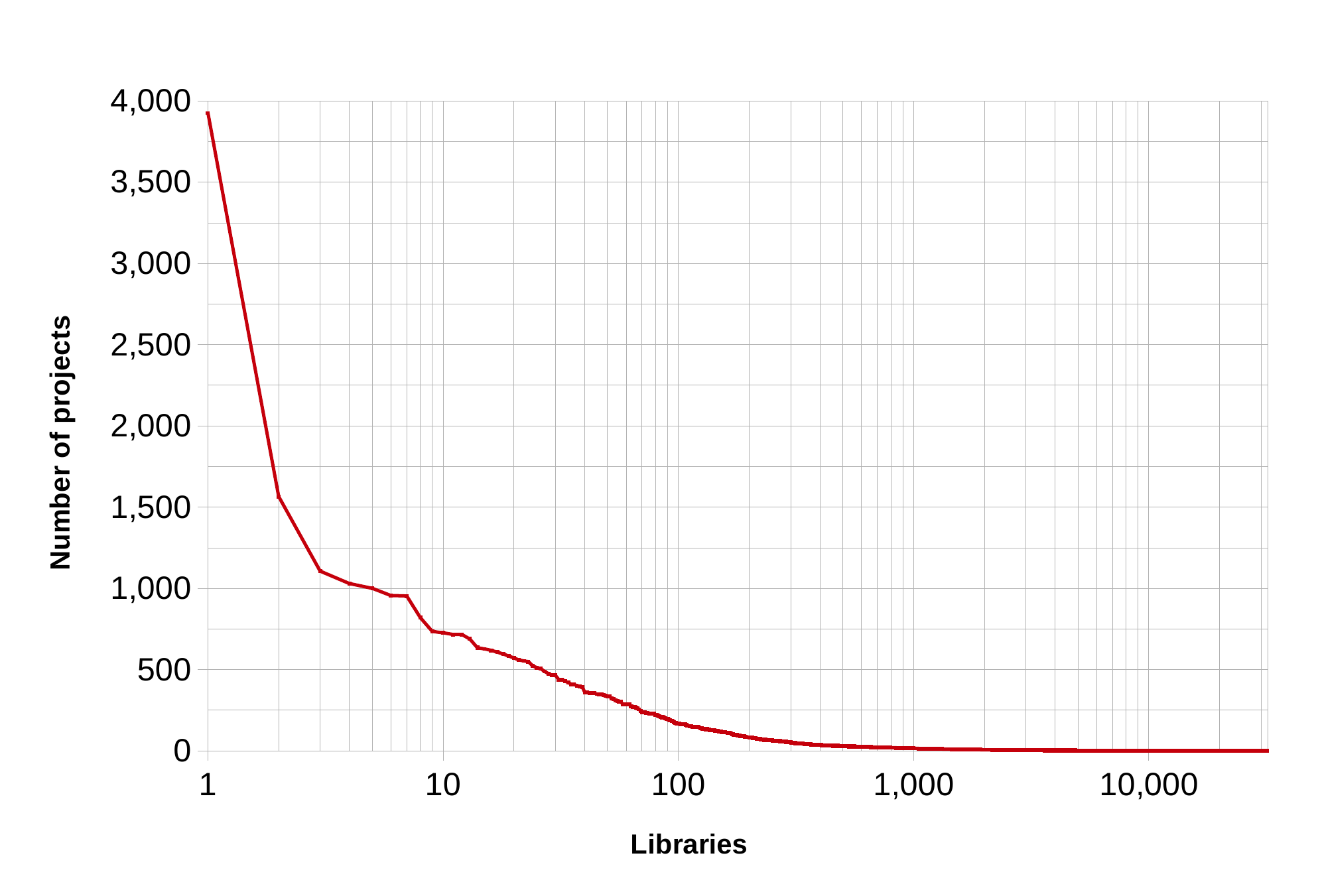}} &     
		\subfigure[\textbf{Dataset DS$_{3}$}]{\label{fig:D3}\includegraphics[width=55mm]{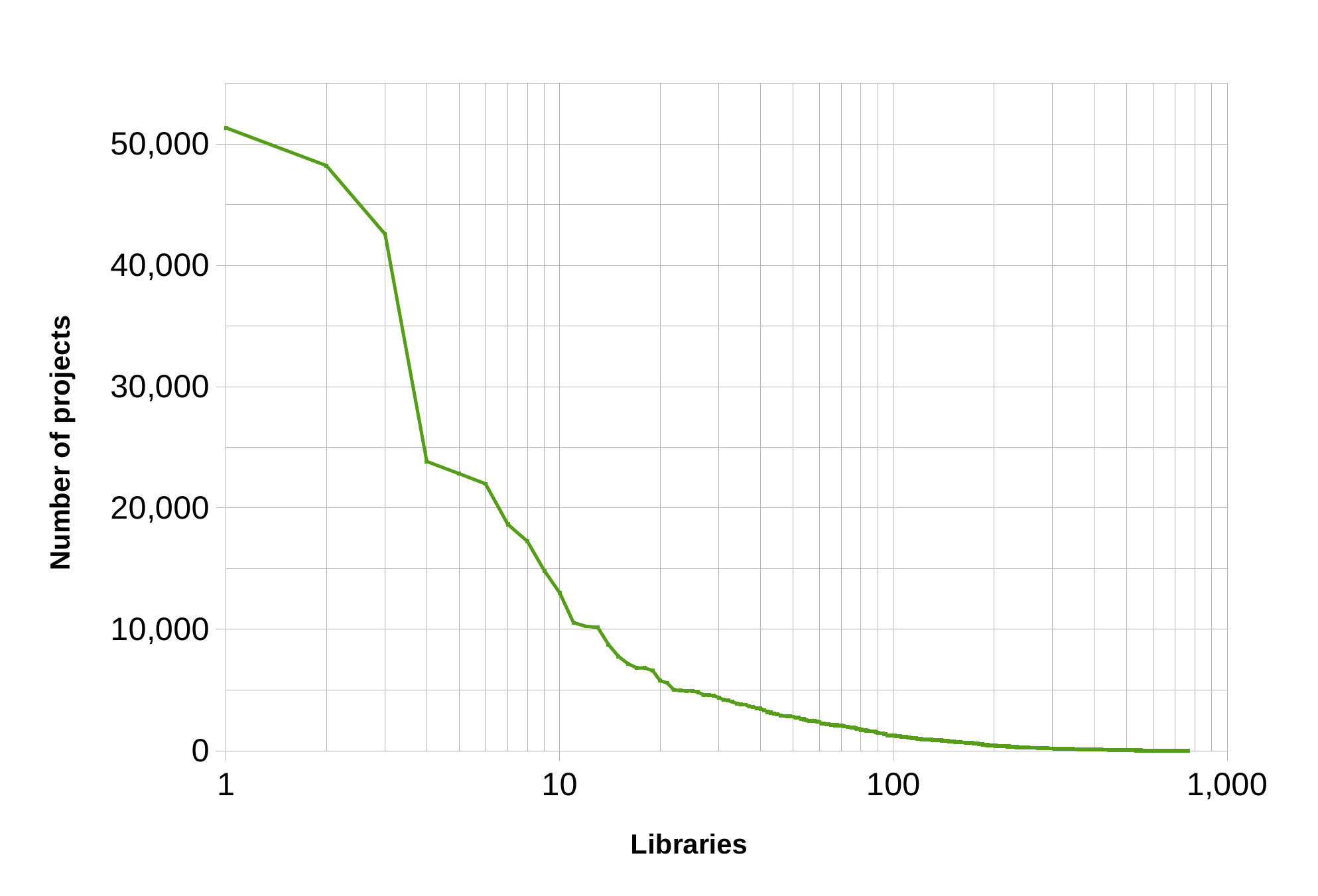}}
	\end{tabular}
	\caption{Frequency of occurrence for the libraries in the three considered datasets.}
	\label{fig:LongTailDatasets}
\end{figure*}



\revised{The bottom of Table~\ref{tab:DatasetsAndSystems} enumerates the considered experimental settings, where the columns represent datasets, and the rows correspond to systems. A cell is filled with a check mark (\faCheck) if the tool in the corresponding row is executed on the dataset in the column, otherwise, it is filled with an uncheck mark (\untick). 
In the ideal case, we should experiment with every possible pair of systems and datasets. This would lead to 12 configurations in total. Indeed, this is computationally expensive, considering the fact that we may spend several hours to execute one configuration. Thus we have to find the most representative configurations by reasoning as follows. DS$_{1}$ was used to evaluate and compare \LR and \CR in the original work~\cite{NGUYEN2019110460}. Instead, DS$_{3}$--originally named MALib\footnote{\url{https://github.com/malibdata/MALib-Dataset}}--was used in the original \LS~\cite{9043686} and \GR~\cite{10.1145/3468264.3468552} papers. Accordingly, in this paper, DS$_{1}$ is used for the evaluation of \LR and \CR, and DS$_{3}$ is for the evaluation of \LS and \GR. Moreover, we curated DS$_{2}$, an extended version of DS$_{1}$ with 4,000 additional projects. DS$_{2}$ is the only dataset applied in the evaluation  of \LR, \CR, \LS, and \GR. 
In this respect, DS$_{2}$ is the common ground to compare the performance of the four systems.}

To give a first idea of how the dataset could cause popularity bias, we counted the number of occurrences for each library in a dataset, and sorted all the libraries in descending order of this number. The frequency for the libraries of DS$_1$, DS$_2$, and DS$_3$ is shown in Fig.~\ref{fig:LongTailDatasets}. As it can be seen, there is always a long tail in all three datasets. For instance, in DS$_1$ the most popular library is \library{junit:junit}, being included in $969$ over $1,200$ projects. Meanwhile, the majority of the TPLs are infrequent \ie we count $10,212$ among $13,497$ libraries appearing only once in the whole dataset. \revised{The datasets reflect how real developers choose TPLs for their systems, either based on their own knowledge/searching on the Web, or possibly, leveraging some recommenders, yet we have no trace of that.}

\revised{
To aim for reliability in the evaluation, we performed the training and testing using different rounds of validation. In particular, we used the ten-fold cross-validation methodology to train and test the systems on the mentioned datasets.}

\subsubsection{Evaluation metrics} \label{sec:Metrics}

Given a set of projects $P$ with a set of libraries $L$, we evaluate 
if the ranked top-$N$ items returned by the 
systems contain long tail items, as well as if they match with the ground-truth data using \emph{Expected popularity complement (EPC)}, \emph{Catalog coverage}, \emph{Precision} and \emph{Recall}. 

\noindent
$\rhd$~\textbf{Expected popularity complement.} The metric measures \emph{novelty}, \ie the ability to provide libraries that are rarely seen, but match with the ground-truth data~\cite{NGUYEN2019110460,10.1145/2487575.2487656,10.1145/2043932.2043955}: 
\begin{equation}\label{eqn:EPC}
	EPC@N = \frac{\sum_{p\in P}\sum_{r=1}^{N} \frac{ rel(p,r)* \left [ 1-pop(REC_{r}(p)) \right ]}{log_{2}(r+1)} }{\sum_{p\in P}\sum_{r=1}^{N} \frac{rel(p,r)}{log_{2}(r+1)}}
\end{equation}
\noindent
where $rel(p,r)$ is 1 if the 
library at the $r$ position of the \emph{top-N} list to project $p$ belongs to the ground-truth data, 0 otherwise; $pop(REC_{r}(p))$ quantifies the popularity of the library at the position $r$ of the 
recommended list. It is the ratio between the number of projects getting $REC_{r}(p)$ as a recommendation, and the number of projects getting the most frequently suggested library as a recommendation, \ie $pop(REC_{r}(p)) = num(REC_{r}(p))/max_{l \in L}(num(l))$. 

\vspace{.1cm}
\noindent
$\rhd$~\textbf{Coverage.} 
This metric gauges \emph{sales diversity}, \ie the ability of a system to suggest a wide range of libraries~\cite{NGUYEN2019110460}, as well as to spread the concentration among all items, rather than focusing only on a limited set of libraries \cite{robillard_recommendation_2014}: $Coverage@N = \left | \cup_{p\in P} REC_{N}(p) \right | / \left | L \right | $. 

\vspace{.05cm}
\noindent 
$\rhd$~\textbf{Precision and Recall.} \emph{Precision} \emph{P@N} is the ratio of 
items matching with the ground-truth data to the total number of provided items; 
\emph{Recall} \emph{R@N} is the ratio of 
matched items 
to the total number of ground-truth items. 



%
%
%
%
%

\subsection{RQ$_3$ Methodology: Re-ranking the recommendations}


Based on xQuAD~\cite{10.1145/1772690.1772780}--a framework for diversifying Web search results--we derive a re-ranking technique to rearrange the recommendation lists, aiming to reduce the number of popular (yet not necessarily useful) libraries, as well as to increase the number of unpopular (but useful) ones in the results. 
Starting from an initial ranked list $R$, we look for the long-tail items in the list, and promote them to upper positions to yield a new list $S$. 
$S$ is initially set to $\varnothing$, and iteratively populated by enrolling new items using the following formula.
\vspace{-.2cm}
\begin{equation} \label{eqn:Eq2}
	\small
	(1-\gamma)\mathcal{P}(l\mid p) +\gamma \mathcal{P}(l, \overline{S}\mid p)
\end{equation}
\noindent where $\gamma$ is the trade-off factor to harmonize accuracy and diversity; $\mathcal{P}(l\mid p)$ is the probability that project $p$ invokes library $l$; $\mathcal{P}(l, \overline{S}\mid p)$ is the probability that $p$ invokes $l$ that does not belong to $S$, computed as follows~\cite{DBLP:conf/flairs/AbdollahpouriBM19}:
\begin{equation}
	\small
	\mathcal{P}(l, \overline{S}\mid p) =  \sum_{c \in H,T} \mathcal{P}(l, \overline{S}\mid c) \mathcal{P}(c\mid p)
\end{equation}
where $H$ and $T$ are the set of head (popular) and tail (long tail) items, respectively. Under the assumption that the libraries are independent of each other with respect to the head and tail categories, $\mathcal{P}(l, \overline{S}\mid c)$ is identified below~\cite{DBLP:conf/flairs/AbdollahpouriBM19}:
\begin{equation} \label{eqn:Eq4}
	\small
	\mathcal{P}(l, \overline{S}\mid c) = \mathcal{P}(c\mid p) \mathcal{P}(\overline{S}\mid c) = \mathcal{P}(c\mid p)\prod_{i \in S}(1- \mathcal{P}(i \mid c,S))
\end{equation}
where $\small \prod_{i \in S}(1- \mathcal{P}(i \mid c,S))$ is set to 1 if item $i$ belongs to both $S$ and category $c$, and $0$ otherwise. 
From Equations~\ref{eqn:Eq2} and~\ref{eqn:Eq4}, the final ranking score for a library $l$ is computed as:
\begin{equation}
	\small
	s_{l} = (1-\gamma) \mathcal{P}(l\mid p) +\gamma \sum_{c \in H,T} \mathcal{P}(c\mid p) \mathcal{P}(l\mid c) \prod_{i \in S}(1- \mathcal{P}(i \mid c,S))
\end{equation}

\revised{The $\gamma$ parameter is used to adjust the trade off between accuracy and fairness~\cite{DBLP:conf/recsys/AbdollahpouriMB19}. By increasing $\gamma$, we give more weight to the second part of Equation~\ref{eqn:Eq2}, \ie improving fairness, but less weight to the first part, \ie decreasing accuracy. In fact, $\gamma$ can be tuned to favor fairness or accuracy. In our experiments, we set $\gamma$ to 0.2 so as to increase fairness, but avoid excessively interfering the accuracy at the same time.}


\section{Results}
\label{sec:Results}

This section reports the study results, addressing the research questions formulated in Section~\ref{sec:Evaluation}.

\subsection{\rqfirst}
\label{sec:RQ1}



The process in Section~\ref{sec:Collection} resulted in a corpus with \numPapers articles from the considered venues in the five most recent years, \ie from 2017 to 2022. We focused on research contributions related to TPL RSSEs and the associated theme, including bias and fairness. From the collected corpus we narrowed down the scope by using three sets of keywords as follows: 
\emph{(i)} \textbf{REC}: ``\emph{recommendation},'' ``\emph{recommender},'' or ``\emph{recommendation systems}''; 
\emph{(ii)} \textbf{LIB}: ``\emph{library},'' ``\emph{libraries},'' ``\emph{TPL},'' ``\emph{TPLs},'' ``\emph{third-party libraries},'' and ``\emph{third-party}''; \emph{(iii)} \textbf{BIF}: ``\emph{bias},'' and ``\emph{fairness}.''
In 
Table~\ref{tab:slr-results} 
we report the number of papers that contain both keywords in the corresponding column and row.\footnote{As the matrix is symmetric, we only list the numbers on the lower left part, and leave the upper right part blank, for the sake of clarity.} Our search targets papers containing one of the following three combinations: either \emph{(i)} REC and LIB; or \emph{(ii)} REC and BIF; or \emph{(iii)} BIF and LIB. 
We mark the associated cells using the green color, and carefully examine these papers. 

\begin{table}[h!]
	\centering
	\footnotesize
	\caption{Number of papers for the related topics.}\label{tab:slr-results}
	\begin{tabular}{|l|c|c|c|}
		\hline
		& REC & LIB & BIF \\ \hline
		REC & 908 & \cellcolor{gray!25} & \cellcolor{gray!25} \\ \hline
		LIB & \cellcolor{green!25}17 & 104 & \cellcolor{gray!25} \\ \hline
		BIF & \cellcolor{green!25}25 & \cellcolor{green!25}2 & 200 \\ \hline
	\end{tabular}
\end{table}


\begin{table*}[t!]
	\centering
	\scriptsize
	\caption{State-of-the-art recommender systems for mining TPLs (Listed in chronological order).}
	\begin{tabular}{|p{1.40cm} | p{1.0cm} | p{0.4cm} | p{1.50cm} | p{5.2cm} |p{5.2cm} | p{0.45cm} |}	\hline
		\textbf{System} & \textbf{Venue} & \textbf{Year} & \textbf{Data source} & \textbf{Working mechanism} & \textbf{Prone to popularity bias?} & \textbf{Avail.}  \\ \hline
		\LR~\cite{LibRec} & WCRE & 2013 & \GH & \LR is built on top of a light collaborative-filtering technique and association mining, looking for libraries that are used by popular projects & The system is exposed to popularity bias by its nature, retrieving only popular libraries thanks to association mining~\cite{10.1145/170035.170072} & \cellcolor{lightgray}\faCheck  \\ \hline
		LibCUP~\cite{SAIED2018164} & JSS & 2017 & \GH  & Usage patterns are discovered by means of DBSCAN~\cite{10.5555/3001460.3001507} -- a hierarchical clustering algorithm & DBSCAN groups libraries that are most frequently co-used by projects. Therefore, popular libraries tend to get recommended more often  & \untick  \\ \hline
		LibFinder~\cite{Ouni:2017:SSL:3032135.3032325} & IST & 2018 & \GH & NSGA-II~\cite{DBLP:journals/tec/DebAPM02} is used to maximize co-usage of libraries, the similarity with the candidates, and the total number of recommended items & 
		A library $L$ can be useful for a system $S$ if $L$ is commonly used with one or more libraries adopted by $S$. Evidence of bias is also reported in the paper & \untick \\ \hline


		\CR~\cite{NGUYEN2019110460} & JSS & 2020 & \GH & \CR employs a collaborative-filtering technique to mine TPLs from similar projects & The system is prone to popularity bias as it recommends libraries coming from projects that are similar & \cellcolor{lightgray}\faCheck  \\ \hline
		Req2Lib~\cite{9054865}& SANER & 2020 & \GH & Using the sequence to sequence technique, Req2Lib learns the library linked-usage information and semantic
		information in natural language & The model is trained with common sequences used by several similar projects, being exposed to popularity bias  & \untick \\ \hline

		\LS~\cite{9043686} & TSE & 2020 & Google Play, \GH, MVN & \LS uses matrix factorization, attempting to neutralize the bias caused by the popularity of TPLs by means of an adaptive weighting mechanism  & \textbf{\emph{Due to its internal design, the system is expected to mitigate the effect of popularity bias}} (We are going to validate this claim in Section~\ref{sec:Results}) & \cellcolor{lightgray}\faCheck \\ \hline
		\GR~\cite{10.1145/3468264.3468552} & ESEC/FSE & 2021 & Google Play  & Built on top of graph neural networks, \GR learns to recommend TPLs through app-library interactions & Thanks to the underlying link prediction technique, \GR is supposed to recommend popular libraries   & \cellcolor{lightgray}\faCheck \\ \hline
	\end{tabular}
	\label{tab:summary}
\end{table*}


Following Table \ref{tab:slr-results}, we notice that 
the combinations of keywords expressed by REC, BIF, and LIB terms occur in 44 
papers. In particular, for the combinations REC and BIF, REC and LIB, and BIF and LIB, we found 25, 17, and 2 articles, respectively. Starting from those, we read the title and abstract to elicit approaches and methodologies, checking whether the papers explicitly deal with 
popularity bias. 
The main findings of the study are discussed in the following.



Among the studies that fall under the combination REC and BIF, Zerouali and Mens \cite{7884645} conducted an empirical study to evaluate the distribution of Java libraries over GitHub projects. Their findings indicated that most of the projects include a limited set of popular libraries that may affect the results of recommender systems. A similar study was carried out on a curated Android dataset to investigate the impact of popular libraries considering several qualitative aspects, \ie malware detection, application repackaging, and static code analysis \cite{LI2019157}. The results showed that common libraries could hamper the performance of repackaging detection tools by introducing both false positives and false negatives. Moreover, the same negative effects were detected in the performance of machine learning-based malware classifiers, \ie removing the most popular libraries increases the overall accuracy on average. He \etal \cite{10.1145/3468264.3468571} examined the long-tail effects during the migration of TPLs employing a fine-grained commit-level analysis of $19,652$ Java GitHub projects. 
The rest of the examined studies covers other types of bias, 
 \ie confirmation bias in testing analysis \cite{salman_controlled_2019}, and fairness in crowd worker recommenders \cite{10.1145/3487571}. \emph{This is actually not related to the topic being considered, \ie popularity bias in TPL recommendation.}



Concerning TPL RSSEs, by combining LIB and REC we ended up with ten papers, however only six of them 
present TPL recommender systems. Apart from these, we searched the same venues to look for similar systems over the same period, and we came across 
\LR~\cite{LibRec} which is among the first TPL RSSEs, being 
used as a baseline for various studies~\cite{9043686,10.1145/3468264.3468552,NGUYEN2019110460}. We included it in our study to further expand the scope of the analysis. 
We report the seven TPL RSSEs 
in Table~\ref{tab:summary}, paying attention to their working mechanisms as well as the possibility of being exposed to popularity bias. 

Overall, the table suggests that diverse underlying techniques are used to recommend libraries. In particular, besides collaborative-filtering based approaches~\cite{NGUYEN2019110460,LibRec}, there are those that employ clustering algorithms~\cite{SAIED2018164}, 
or NSGA-II~\cite{Ouni:2017:SSL:3032135.3032325}. Notably, deep neural networks~\cite{9043686,10.1145/3468264.3468552,9054865} and matrix factorization~\cite{9043686} also found their application in TPL recommendations. \LR works on top of a light collaborative-filtering technique and association mining, retrieving libraries that are used by popular projects. LibCUP~\cite{SAIED2018164} mines usage patterns using DBSCAN, a hierarchical clustering algorithm. LibFinder \cite{Ouni:2017:SSL:3032135.3032325} makes use of the NSGA-II (\revised{Non-dominated Sorting Genetic Algorithm}) multi-objective search algorithm to perform recommendations. \CR~\cite{NGUYEN2019110460} exploits a graph-based structure to recommend relevant TPLs given the developer's context. 
Req2Lib \cite{9054865} suggests relevant TPLs starting from the textual description of the requirements to handle the cold-start problem by combining a Sequence-to-Sequence network with a doc2vec pre-trained model. 
Similarly, GRec \cite{10.1145/3468264.3468552} encodes mobile apps, TPLs, and their interactions in an app-library graph. Afterward, it uses a graph neural network to distill the relevant features to increase the overall accuracy. Altogether, \emph{all these systems are not conceived to mitigate the effect of popularity bias}. 


Apart from the seven studies presenting TPL RSSEs previously described, the remaining four of the eleven belonging to the combination of LIB and REC, tackle different issues as reported as follows. Chen \etal \cite{8630054} proposed an unsupervised deep learning approach to embed both usage and description semantics of TPLs to infer mappings at the API level. 
An approach~\cite{10.1007/s10664-018-9657-y} based on Stack Overflow was proposed to recommend analogical libraries, \ie a library similar to the ones that developers already use. Nafi \etal \cite{9825781} developed XLibRec, a tool that recommends analogical libraries across different programming languages. 
Rubei \etal \cite{9825861} investigated the usage of a learning-to-rank mechanism to embody explicit user feedback in TPLs recommenders. In summary, \emph{there is no paper among the ones discussed above copes with popularity bias in RSSEs.}

\begin{figure*}[h!]
	\centering    
	\begin{tabular}{c c c c}		
		\subfigure[\textbf{DS$_1$}, N=10]{\label{fig:LibRecD1_10}\includegraphics[width=32mm]{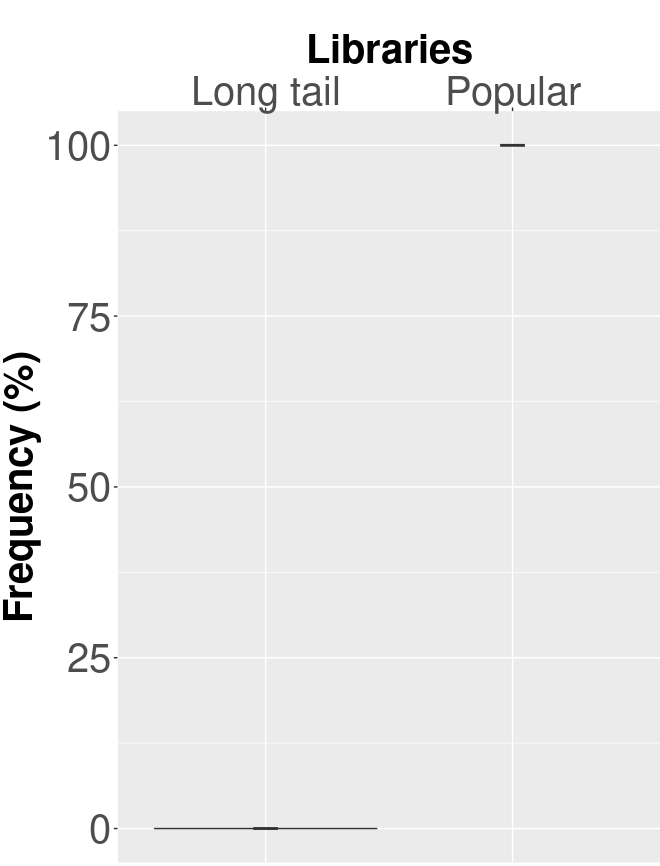}} &		
		\subfigure[\textbf{DS$_1$}, N=20]{\label{fig:LibRecD1_20}\includegraphics[width=32mm]{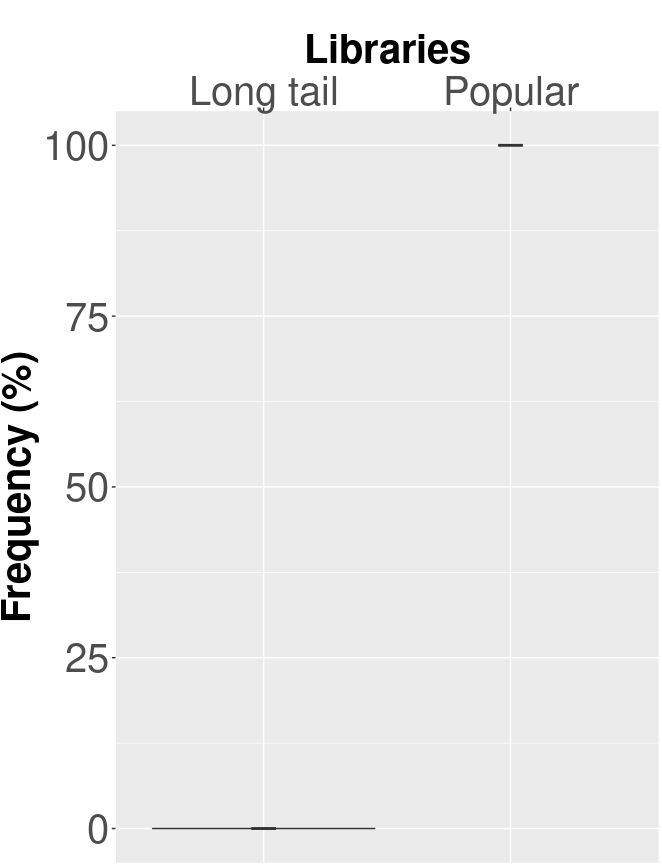}} &	
		\subfigure[\textbf{DS$_2$}, N=10]{\label{fig:LibRecD2_10}\includegraphics[width=32mm]{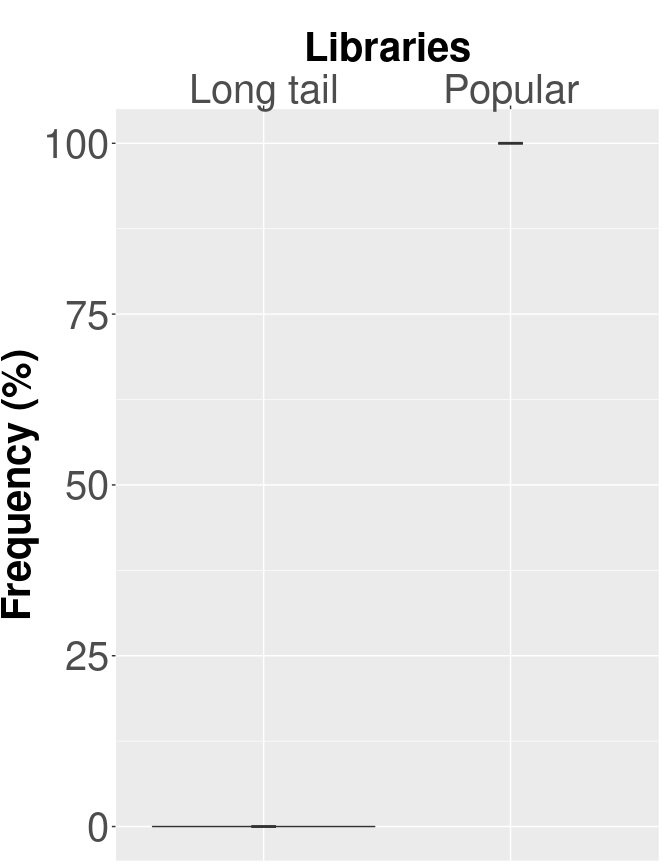}} &		
		\subfigure[\textbf{DS$_2$}, N=20]{\label{fig:LibRecD2_20}\includegraphics[width=32mm]{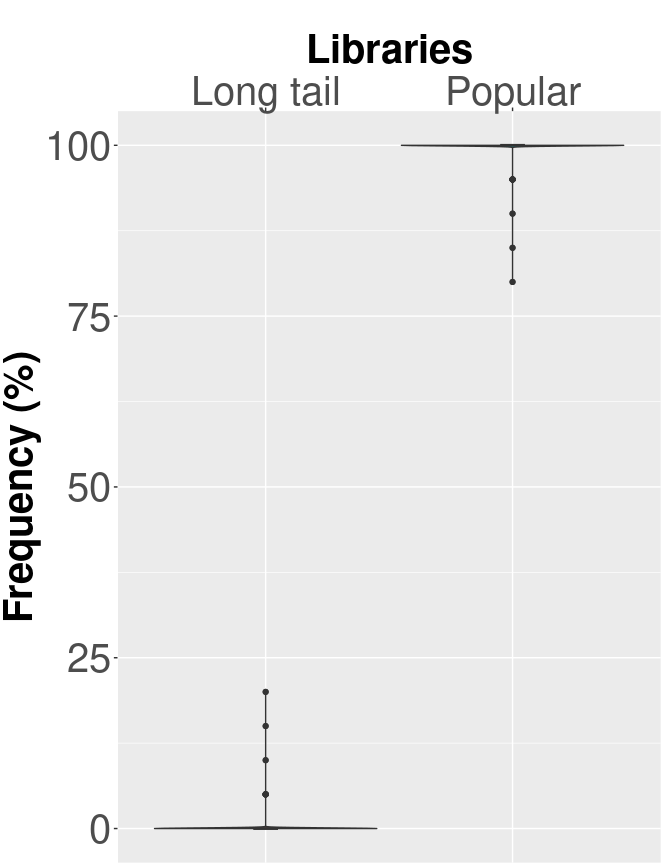}} 	
	\end{tabular}
	\caption{\LR: Recommendation of popular and long tail libraries.}
	\label{fig:ResultsLibRec}
\end{figure*}

\begin{figure*}[h!]
	\centering    
	\begin{tabular}{c c c c}		
		\subfigure[\textbf{DS$_1$}, N=10]{\label{fig:CrossRecD1_10}\includegraphics[width=32mm]{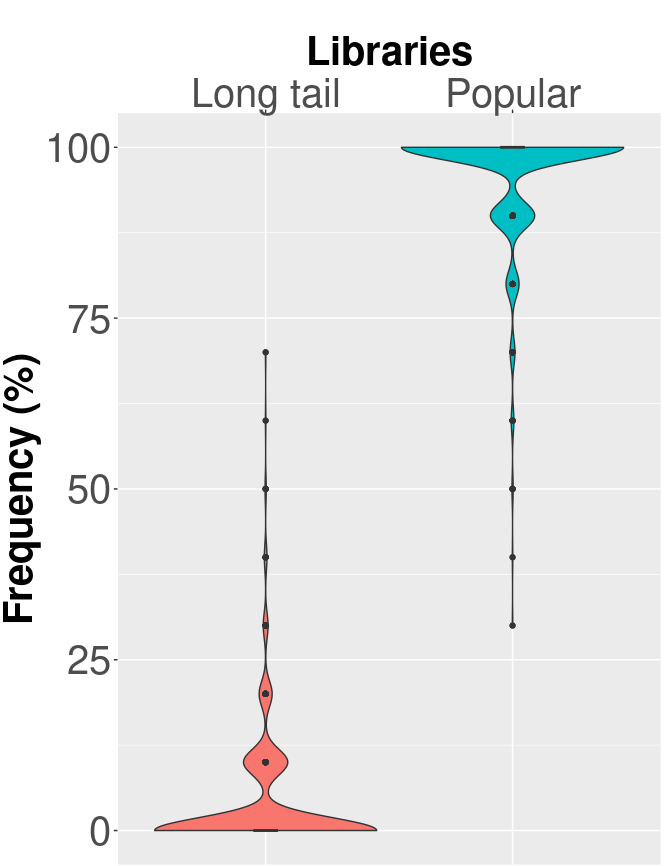}} &		
		\subfigure[\textbf{DS$_1$}, N=20]{\label{fig:CrossRecD1_20}\includegraphics[width=32mm]{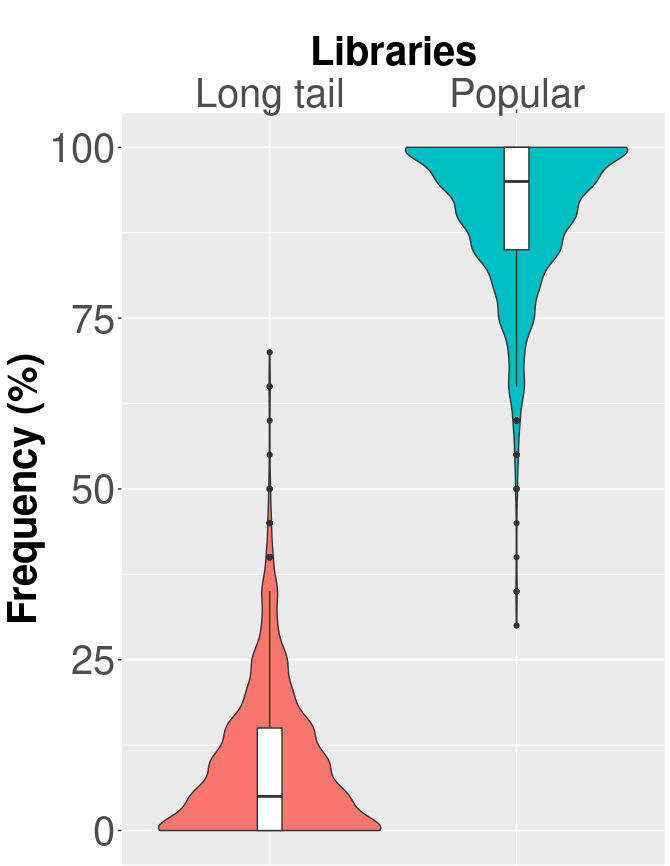}} &	
		\subfigure[\textbf{DS$_2$}, N=10]{\label{fig:CrossRecD2_10}\includegraphics[width=32mm]{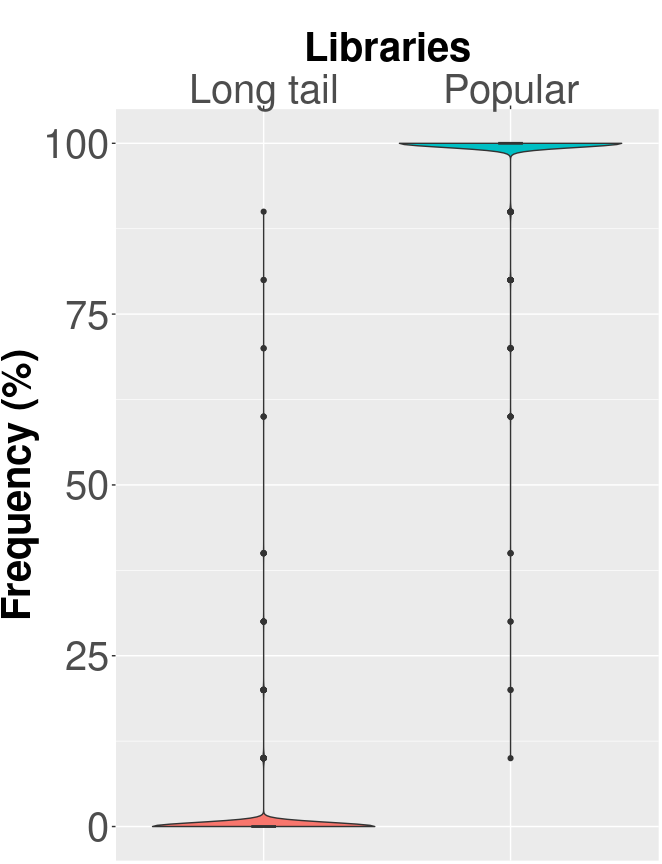}} &		
		\subfigure[\textbf{DS$_2$}, N=20]{\label{fig:CrossRecD2_20}\includegraphics[width=32mm]{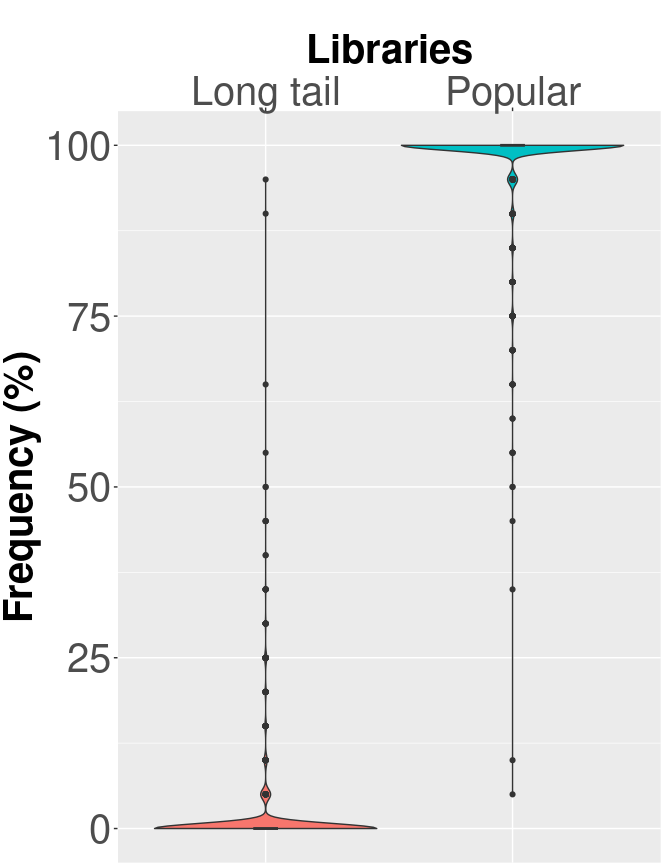}}	
	\end{tabular}
	\caption{\CR: Recommendation of popular and long tail libraries.}
	\label{fig:ResultsCrossRec}
\end{figure*}

\begin{figure*}[h!]
	\centering    
	\begin{tabular}{c c c c}		
		\subfigure[\textbf{DS$_2$}, N=5]{\label{fig:LibSeekD2_5}\includegraphics[width=32mm]{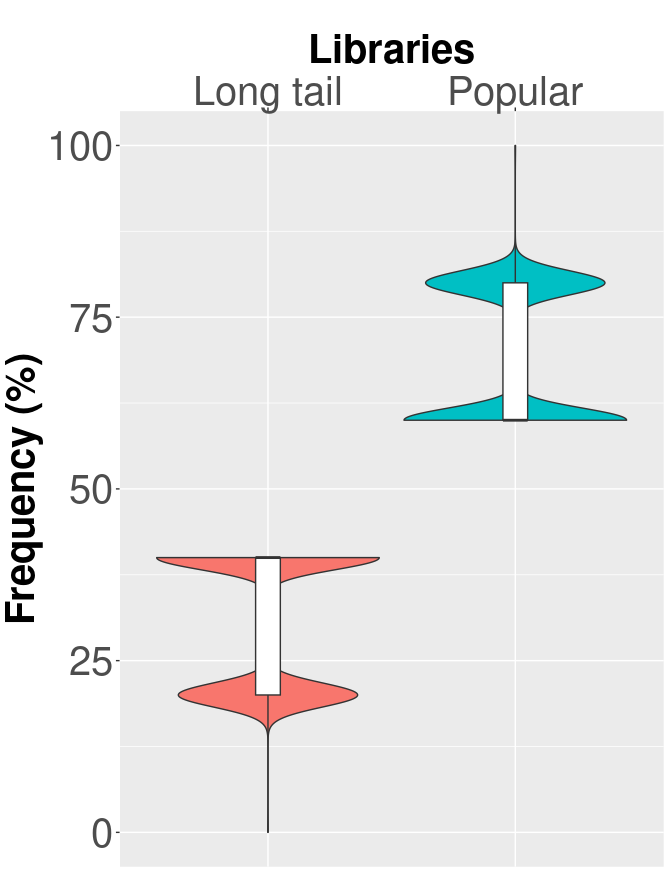}} &		
		\subfigure[\textbf{DS$_2$}, N=10]{\label{fig:LibSeekD2_10}\includegraphics[width=32mm]{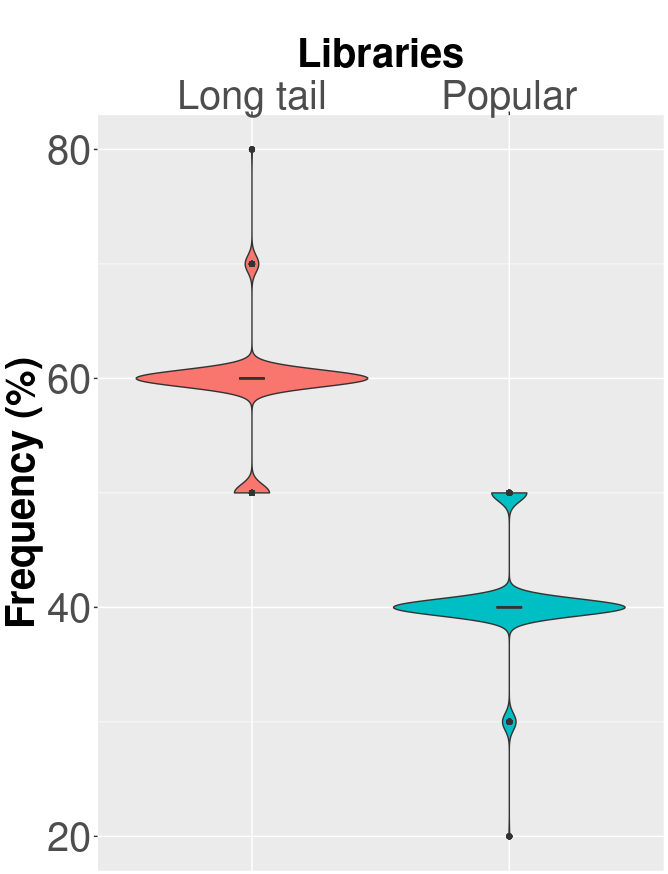}} &	
		\subfigure[\textbf{DS$_3$}, N=5]{\label{fig:LibSeekD3_5}\includegraphics[width=32mm]{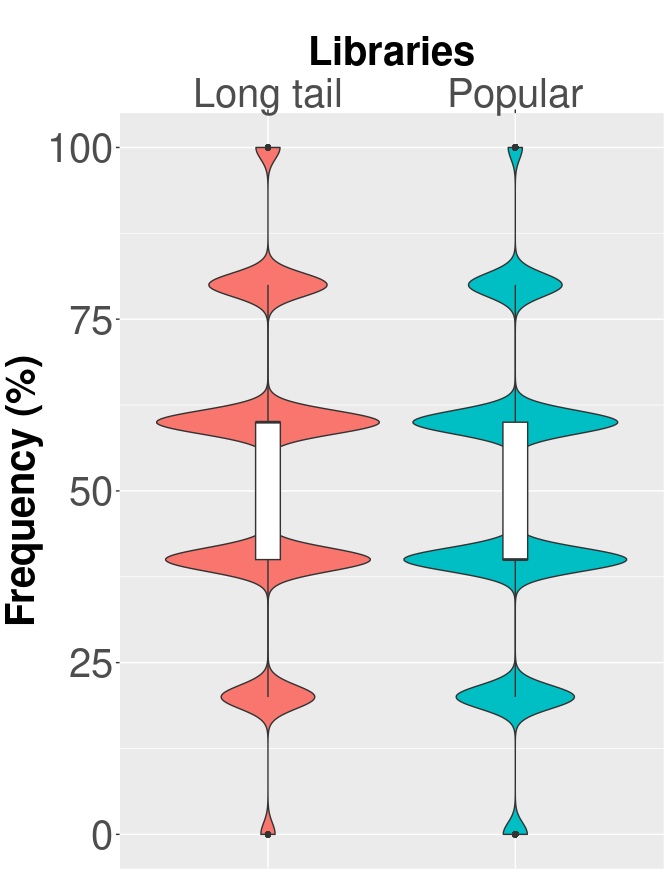}} &		
		\subfigure[\textbf{DS$_3$}, N=10]{\label{fig:LibSeekD3_10}\includegraphics[width=32mm]{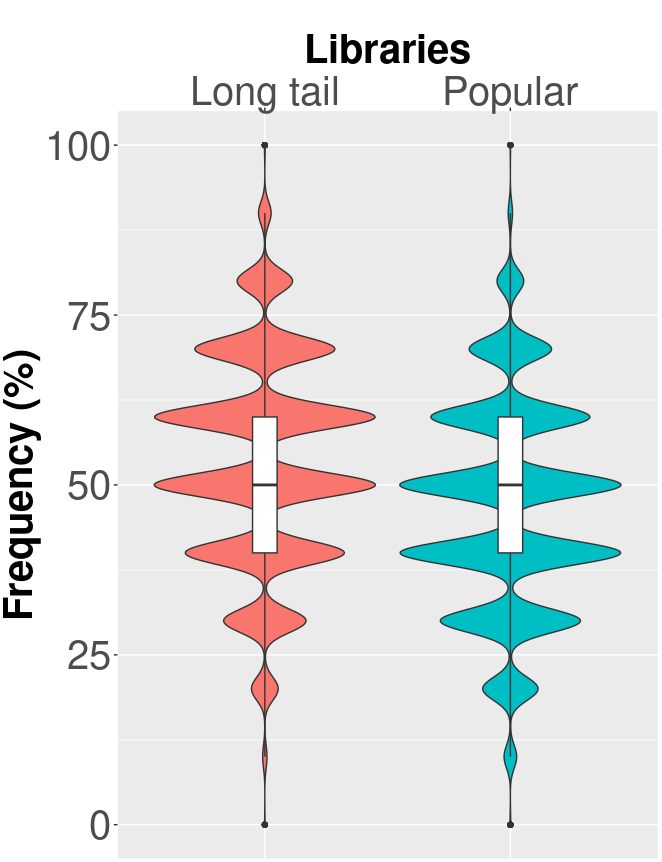}}	  
	\end{tabular}
	\caption{\LS: Recommendation of popular and long tail libraries.}
	\label{fig:ResultsLibSeek}
\end{figure*}

\begin{figure*}[h!]
	\centering    
	\begin{tabular}{c c c c}		
		\subfigure[\textbf{DS$_2$}, N=5]{\label{fig:GRecD2_5}\includegraphics[width=32mm]{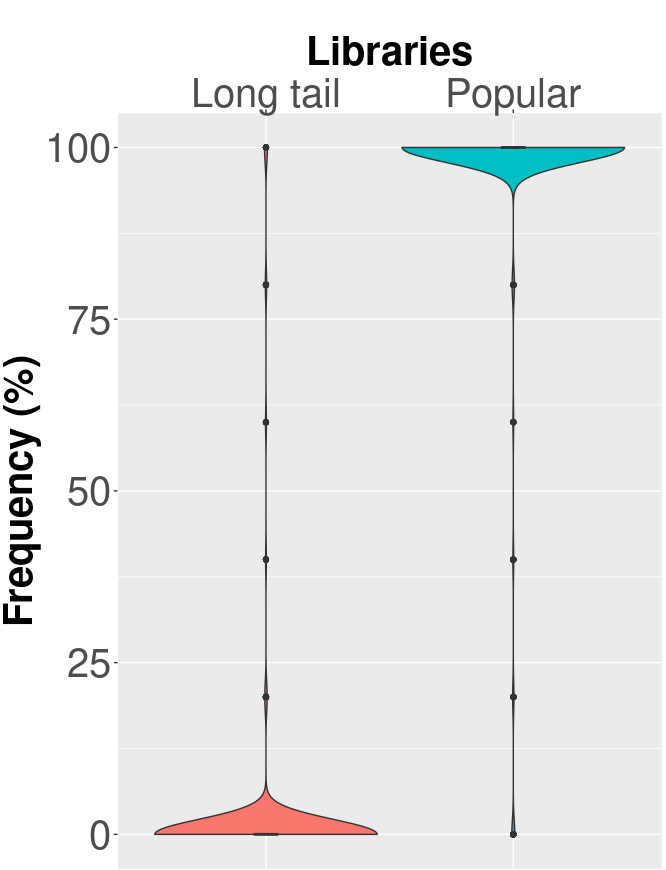}} &		
		\subfigure[\textbf{DS$_2$}, N=10]{\label{fig:GRecD2_10}\includegraphics[width=32mm]{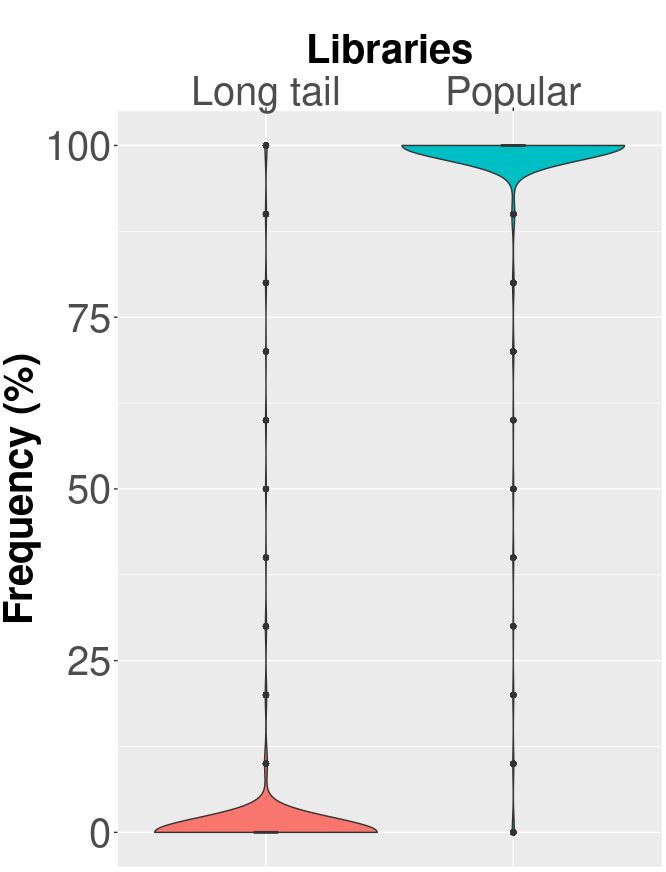}} &	
		\subfigure[\textbf{DS$_3$}, N=5]{\label{fig:GRecD3_5}\includegraphics[width=32mm]{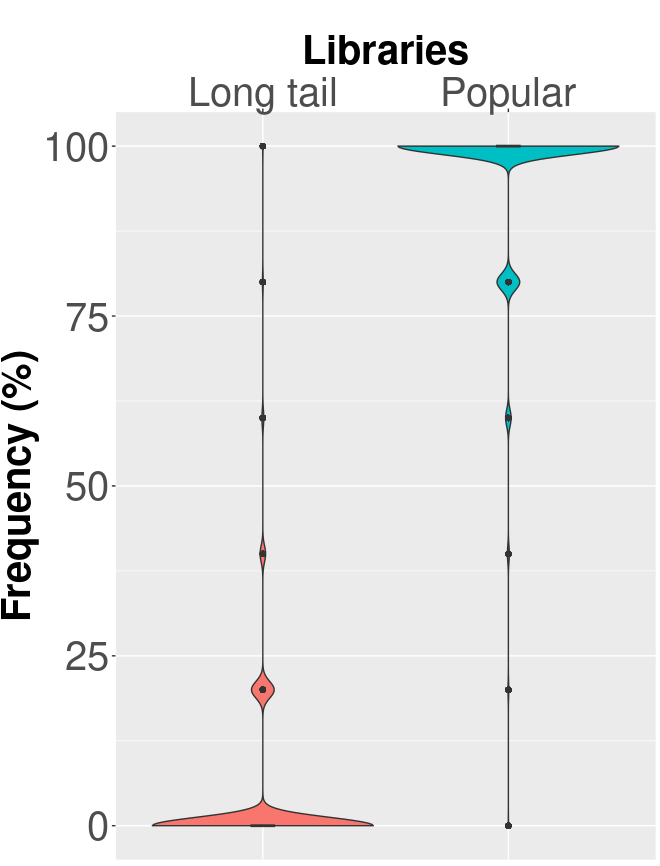}} &		
		\subfigure[\textbf{DS$_3$}, N=10]{\label{fig:GRecD3_10}\includegraphics[width=32mm]{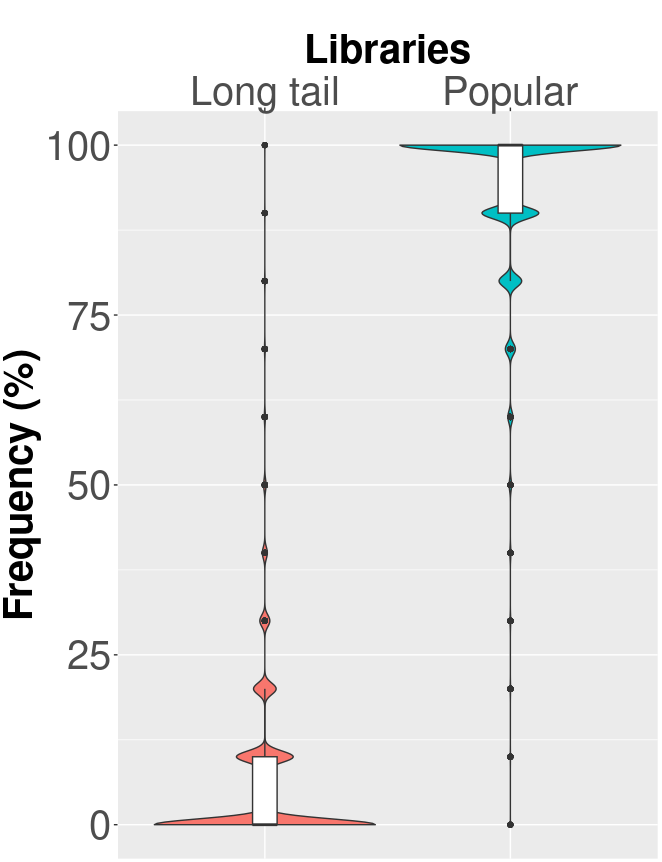}}	  
	\end{tabular}
	\caption{\GR: Recommendation of popular and long tail libraries.}
	\label{fig:ResultsGRec}
\end{figure*}

He \etal~\cite{9043686} are, to the best of our knowledge, the first to address the issue of recommending diversified libraries. They found out that the prediction results provided by existing TPL RSSEs tend to favor a small set of libraries. 
They also highlighted that recommending popular TPLs hampers the usefulness of the recommendations, resulting in the lack of novelty and serendipity, which are deemed to be among the desired features of RSSEs~\cite{NGUYEN2019110460}. Thus, He \etal proposed \LS as the first system to provide diversified libraries. The system uses an adaptive weighting mechanism to neutralize popularity bias by assigning higher weights to less popular TPLs. The evaluation conducted on a curated Android dataset confirmed that \emph{\LS succeeds in providing a wide range of libraries, thus increasing novelty in the recommendation outcomes}. 

\vspace{.1cm}
\begin{shadedbox}
	\small{\textbf{Answer to RQ$_1$.} So far, dealing with popularity bias in TPL recommender systems has not received adequate attention from the Software Engineering community. Among the considered studies, \LS is the sole attempt to improve diversity in the recommendation results.} 
\end{shadedbox}

\subsection{\rqsecond}
\label{sec:RQ2}

As explained in Section~\ref{sec:RQ2Methodology}, \numSys representative systems were chosen, 
\ie \LR~\cite{LibRec}, \CR~\cite{NGUYEN2019110460}, \LS~\cite{9043686}, and \GR~\cite{10.1145/3468264.3468552}, whose source implementation is available (marked with a~\faCheck~in Table~\ref{tab:summary}), allowing us to experiment on real-world datasets according to our needs. Following existing studies~\cite{DBLP:conf/flairs/AbdollahpouriBM19,DBLP:conf/recsys/AbdollahpouriMB19}, we split each list in Fig.~\ref{fig:LongTailDatasets} into two parts, the first $20\%$ TPLs are determined as popular, and the remaining $80\%$ libraries represent the long tail. For each project, we count in the recommendations the percentage of TPLs that belong to the popular part and the long tail. Moreover, we consider another parameter that affects the final recommendations, \ie N, the cut-off value of the ranked list. As \LR and \CR return a long list of items, we consider two cut-off values, \ie N=$\{10,20\}$. Instead, \LS and \GR only provide short ranked lists, and thus only small cut-off values are allowed, \ie N=$\{5,10\}$. Besides popularity, we measure the accuracy using precision and recall, two popular metrics for evaluating RSSEs~\cite{10.1145/3468264.3468552,NGUYEN2019110460}. The evaluation is focused on two aspects, \ie if the systems \emph{(i)} are subject to popularity bias; and \emph{(ii)} can provide accurate recommendations. 

\subsubsection{Popularity bias}

Based on the obtained results, 
we study how the \numSys systems deal with popularity bias 
as follows.

\vspace{.05cm}
\noindent
$\rhd$~\textbf{\LR.} The results obtained with \LR on DS$_1$ and DS$_2$ are depicted in Fig.~\ref{fig:ResultsLibRec}. It is evident that \LR is severely exposed to popular bias, \ie the entire plots concentrate on the 100\% level of popularity, implying that all the TPLs belong to the 20\% frequent items. Essentially, \LR barely provides any TPL in the long tail. This holds for both datasets and cut-off values. This happens due to the fact that \LR works based on association rule mining~\cite{10.1145/170035.170072}, \ie \emph{recommending the items widely used by the majority of projects.}

\vspace{.05cm}
\noindent
$\rhd$~\textbf{\CR.} Fig.~\ref{fig:ResultsCrossRec} shows that, compared to \LR, \CR can cope with popularity bias better. On DS$_1$, when N=$10$, \CR generally recommends very popular libraries. However, it can provide few libraries in the long tail. The ability to recommend rare TPLs is improved when a longer list is considered as shown in Fig.~\ref{fig:CrossRecD1_20}, when N=$20$. The system provides more popular libraries and fewer unpopular libraries by DS$_2$ as we can see in Fig.~\ref{fig:CrossRecD2_10} and Fig.~\ref{fig:CrossRecD2_20}. This means \emph{\CR is less successful in defusing popularity bias on datasets such as DS$_2$, where there are more TPLs, as it can be noticed from the long tail.}

\noindent
$\rhd$~\textbf{\LS.} As shown in Fig.~\ref{fig:LibSeekD2_5}, \LS suffers from popularity bias on DS$_2$ when N=$5$. Nevertheless, it improves fairness by recommending TPLs in the long tail when a longer ranked list is considered, \ie N=$10$. In particular, Fig.~\ref{fig:LibSeekD2_10} shows that \LS introduces more infrequent items compared to the popular ones. By comparing Fig.~\ref{fig:LibSeekD2_5} and Fig.~\ref{fig:LibSeekD2_10}, we conclude that \LS provides long tail TPLs late in the list. The results in Fig.~\ref{fig:LibSeekD3_5}, and Fig.~\ref{fig:LibSeekD3_10} confirm the claim made by the authors of \LS~\cite{9043686} (and thus the analysis in Table~\ref{tab:summary}), \ie the tool can handle well the items in the long tail, mitigating the effect of popularity bias. Notably, the ability of \LS to provide long tail items is evident only in DS$_3$, which contains a large number of projects, but a small number of TPLs (see Table~\ref{tab:DatasetsAndSystems}). Meanwhile, by DS$_2$, where there are fewer projects but more libraries, the system fails to cope with long tail items as shown in Fig.~\ref{fig:LibSeekD2_5}. Altogether, this shows that \emph{while being able to cope with popularity bias in general, \LS does not succeed in some certain cases. Indeed, the characteristics of a dataset might be a contributing factor in the ability of \LS to mitigate such a bias.}

\vspace{.05cm}
\noindent
$\rhd$~\textbf{\GR.} As shown in Fig.~\ref{fig:ResultsGRec}, the system mainly recommends popular TPLs. By both DS$_2$ (Fig.~\ref{fig:GRecD2_5} and Fig.~\ref{fig:GRecD2_10}) and DS$_3$ (Fig.~\ref{fig:GRecD3_5} and Fig.~\ref{fig:GRecD3_10}) as well as both values of N, most of the TPLs provided by \GR are frequent, \ie the plots are concentrated on the maximum level of popularity, only tiny fractions of the libraries reside in the long tail. This means that it provides to developers only libraries that are used by several projects, 
while ignoring the less popular ones. 
In a nutshell, the results in Fig.~\ref{fig:ResultsGRec} reveal that 
\emph{\GR cannot deal with popularity bias in the recommendation results.}

\begin{table}[h!]
	\centering
	\scriptsize	
	\caption{Prediction accuracy.}
	\begin{tabular}{|p{1.20cm} |p{1.20cm} | p{1.80cm} | p{1.80cm} |}	\hline
		\textbf{Systems} & \textbf{Datasets} & Mean Precision & Mean Recall \\ \hline
		\multirow{2}{*}{\LR} & DS$_1$	 & 0.211 & 0.217 \\ \cline{2-4}
		& \cellcolor{lightgray}DS$_2$	 & \cellcolor{lightgray}0.272 & \cellcolor{lightgray}0.243 \\ \hline
		\multirow{2}{*}{\CR} & DS$_1$	 & 0.289  & 0.123 \\ \cline{2-4}
		& \cellcolor{lightgray}DS$_2$	 & \cellcolor{lightgray}\textbf{0.342} & \cellcolor{lightgray}\textbf{0.328} \\ \hline
		\multirow{2}{*}{\LS} & \cellcolor{lightgray}DS$_2$	 & \cellcolor{lightgray}0.031  & \cellcolor{lightgray}0.090 \\ \cline{2-4}
		& DS$_3$	 & 0.211 & 0.694 \\ \hline
		\multirow{2}{*}{\GR} & \cellcolor{lightgray}DS$_2$	 & \cellcolor{lightgray}0.062 & \cellcolor{lightgray}0.212 \\ \cline{2-4}
		& DS$_3$  & 0.221 & 0.713 \\ \hline
	\end{tabular}
	\label{tab:PredictionAccuracy}
\end{table}

%

\subsubsection{Accuracy}

An important feature of any recommender system is its ability to provide accurate items~\cite{robillard_recommendation_2014}. We analyze the prediction 
of the considered systems by referring to Table~\ref{tab:PredictionAccuracy}. Concerning the results obtained on DS$_2$ (rows marked in gray), it can be noticed that, although \LR and \CR are not very successful at providing diversified TPLs (see Fig.~\ref{fig:ResultsLibRec} and Fig.~\ref{fig:ResultsCrossRec}), they obtain a better prediction accuracy compared to that of \LS and \GR. On DS$_2$, \CR gets $0.342$ and $0.328$ as precision and recall, being the best tool concerning prediction accuracy. \LS is fairly good at coping with popularity bias; however, this comes at a price: It achieves a very low accuracy compared to the others. The precision and recall scores for DS$_2$ are $0.031$ and $0.090$, respectively, and for DS$_3$ $0.211$ and $0.694$, respectively. 
Likewise, \GR also fails to provide 
relevant items, achieving a low accuracy on DS$_2$, \ie precision and recall are 0.062 and 0.212, respectively. Only by DS$_3$, \GR improves its prediction 
by getting $0.221$ and $0.713$ as precision and recall.



\vspace{.1cm}
\begin{shadedbox}
	\small{\textbf{Answer to RQ$_2$.} Three among the 
		systems recommend very popular third-party libraries, while usually ignoring those that are rarely used by projects. Only \LS is able to recommend libraries in the long tail, however, it fails to maintain a trade-off between fairness and precision, resulting in a low accuracy.}
\end{shadedbox}

\subsection{\rqthird}
\label{sec:RQ3}
\begin{figure*}[t!]
	\centering    
	\begin{tabular}{c c c}					
		\subfigure[Expected popularity complement]{\label{fig:EPC_LibRec_DS2}\includegraphics[width=56mm]{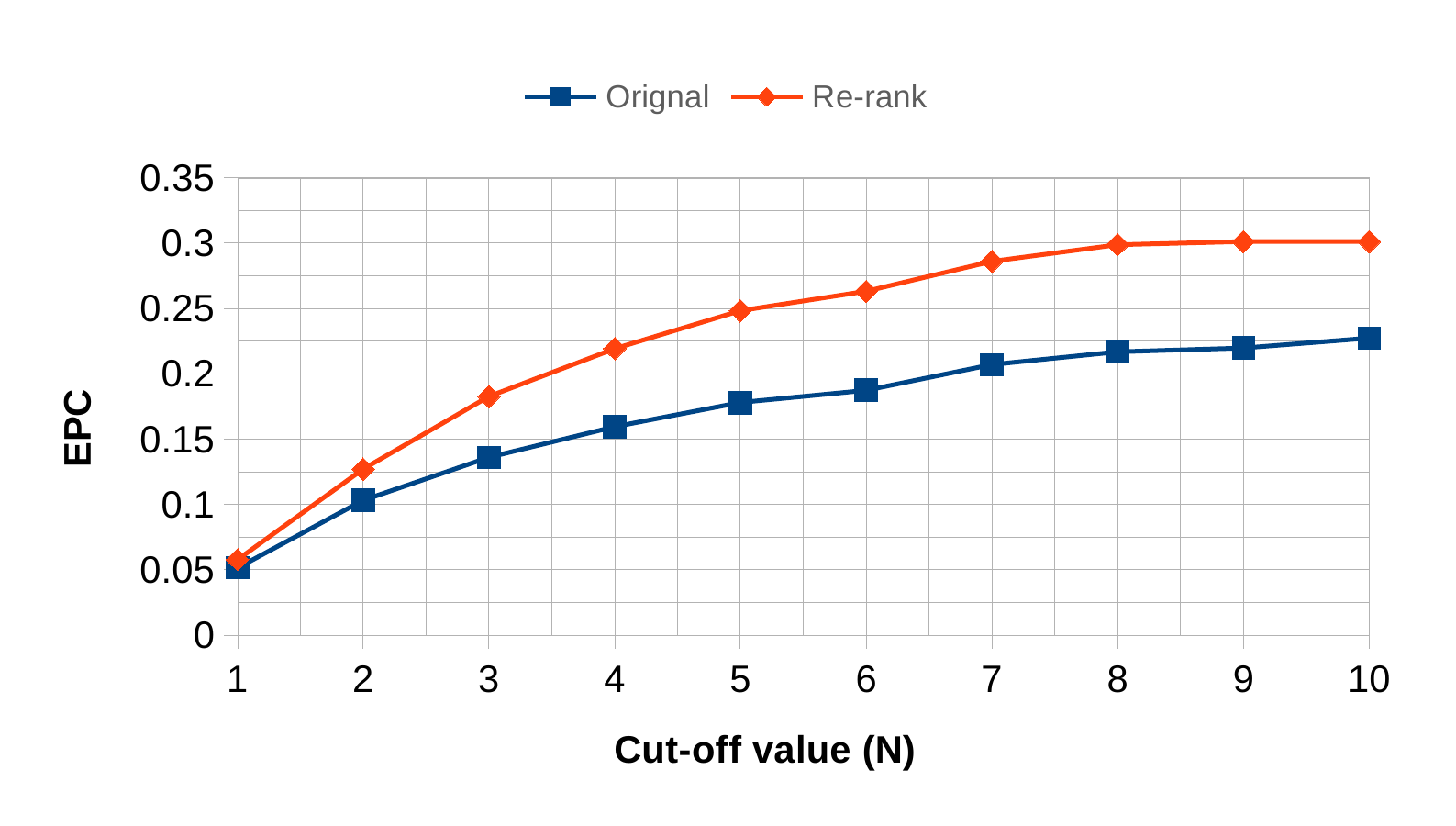}} &		
		\subfigure[Coverage]{\label{fig:Catalog_LibRec_DS2}\includegraphics[width=56mm]{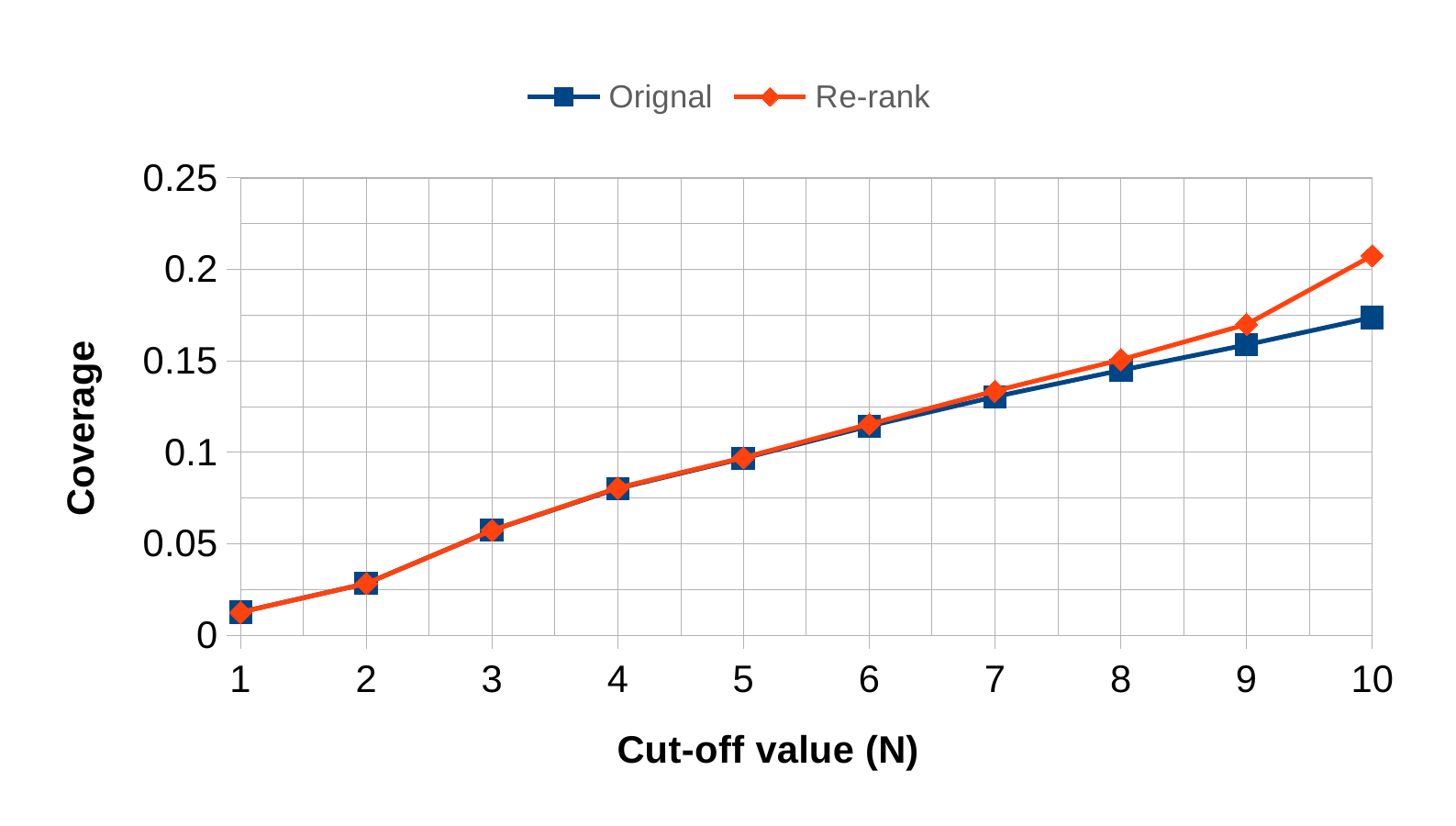}} &
		\subfigure[Precision and recall]{\label{fig:PrecisionRecall_LibRec_DS2}\includegraphics[width=56mm]{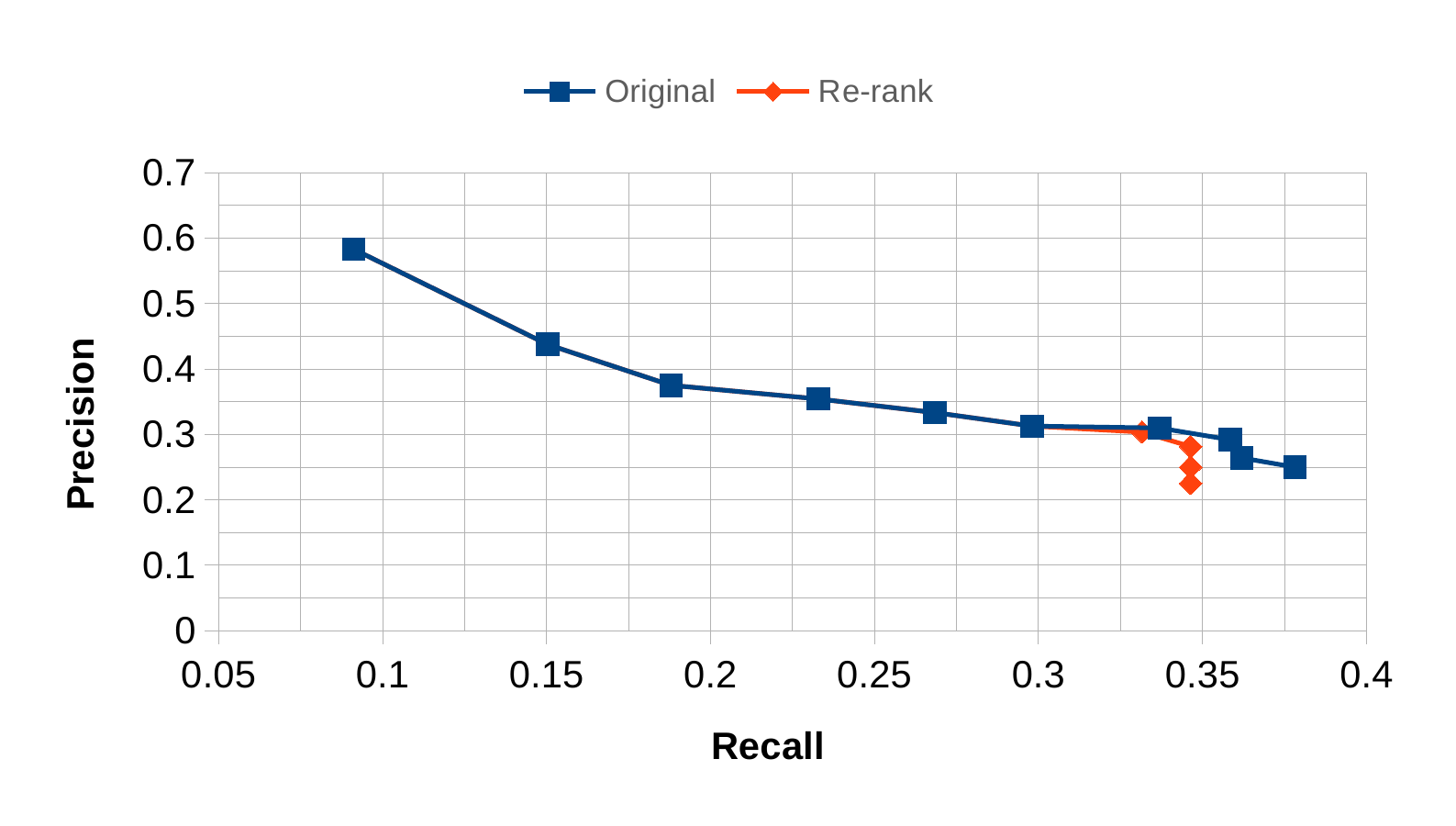}} 
	\end{tabular}
	\caption{Results obtained on the recommendations by running \LR with DS$_2$.}
	\label{fig:LibRecEvalutionMetrics}
\end{figure*}

\begin{figure*}[t!]
	\centering    
	\vspace{-.2cm}
	\begin{tabular}{c c c}			
		\subfigure[Expected popularity complement]{\label{fig:EPC_CrossRec_DS1}\includegraphics[width=56mm]{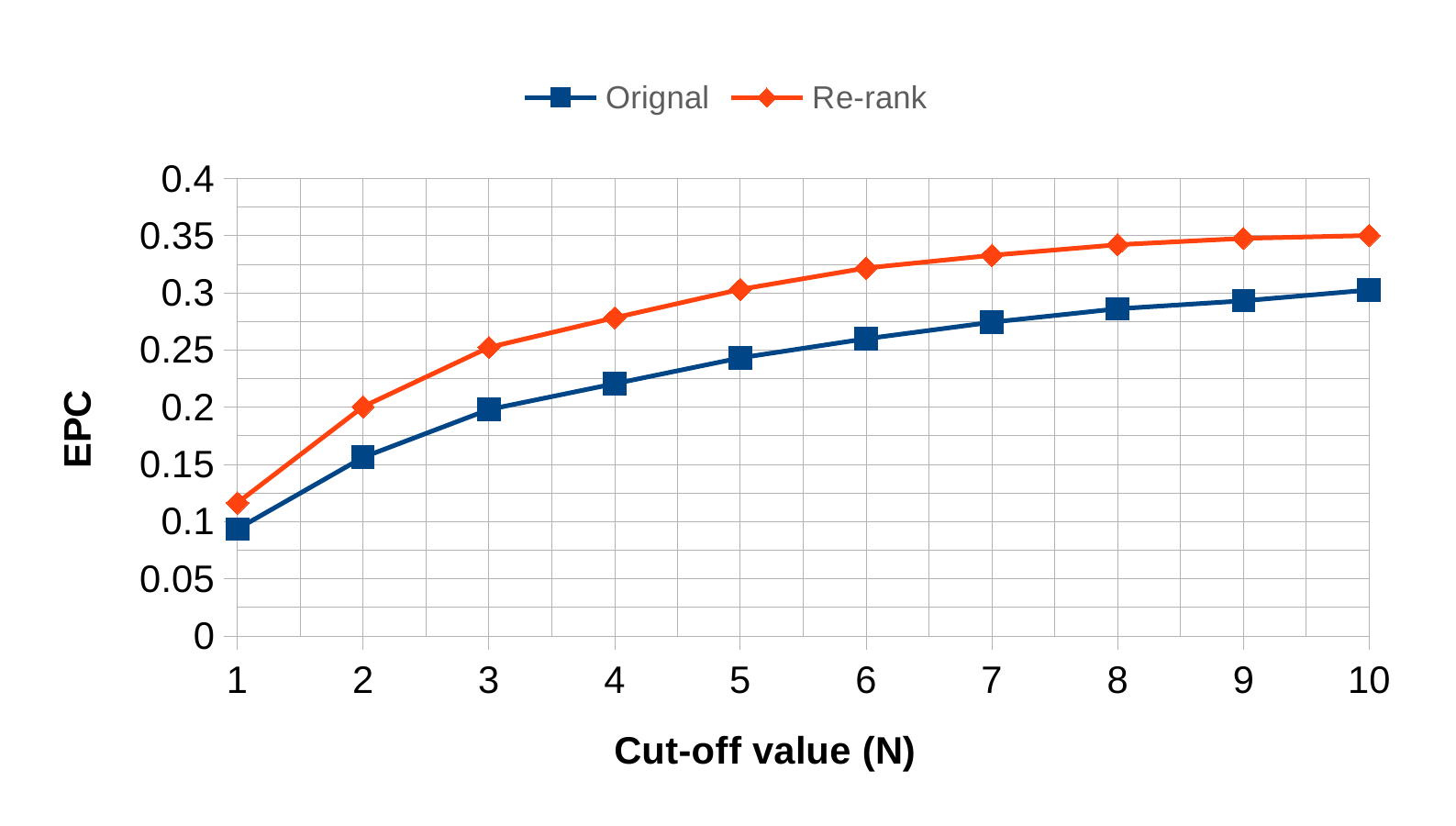}} &		
		\subfigure[Coverage]{\label{fig:Catalog_CrossRec_DS1}\includegraphics[width=56mm]{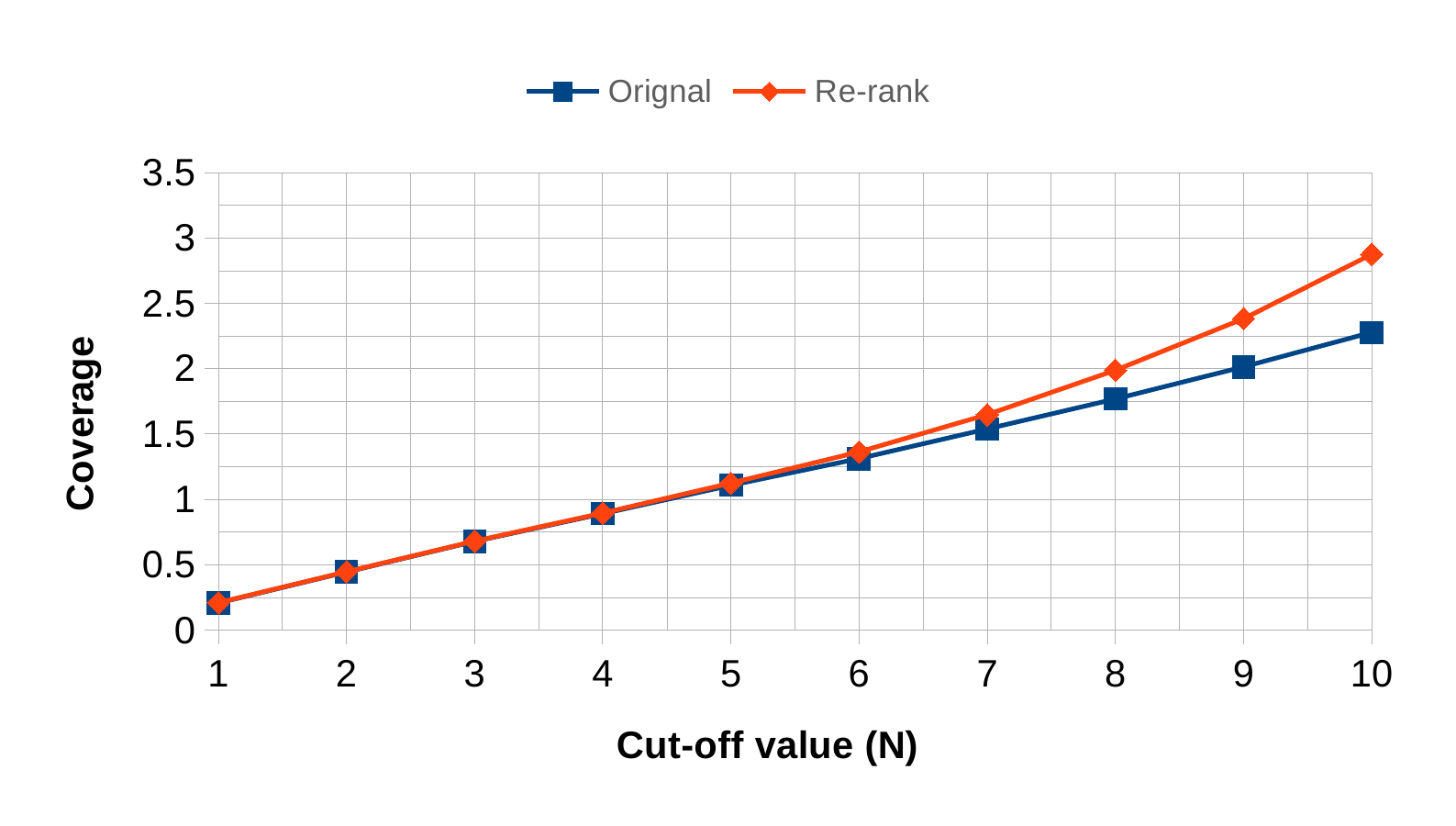}} &
		\subfigure[Precision and recall]{\label{fig:PrecisionRecall_CrossRec_DS1}\includegraphics[width=56mm]{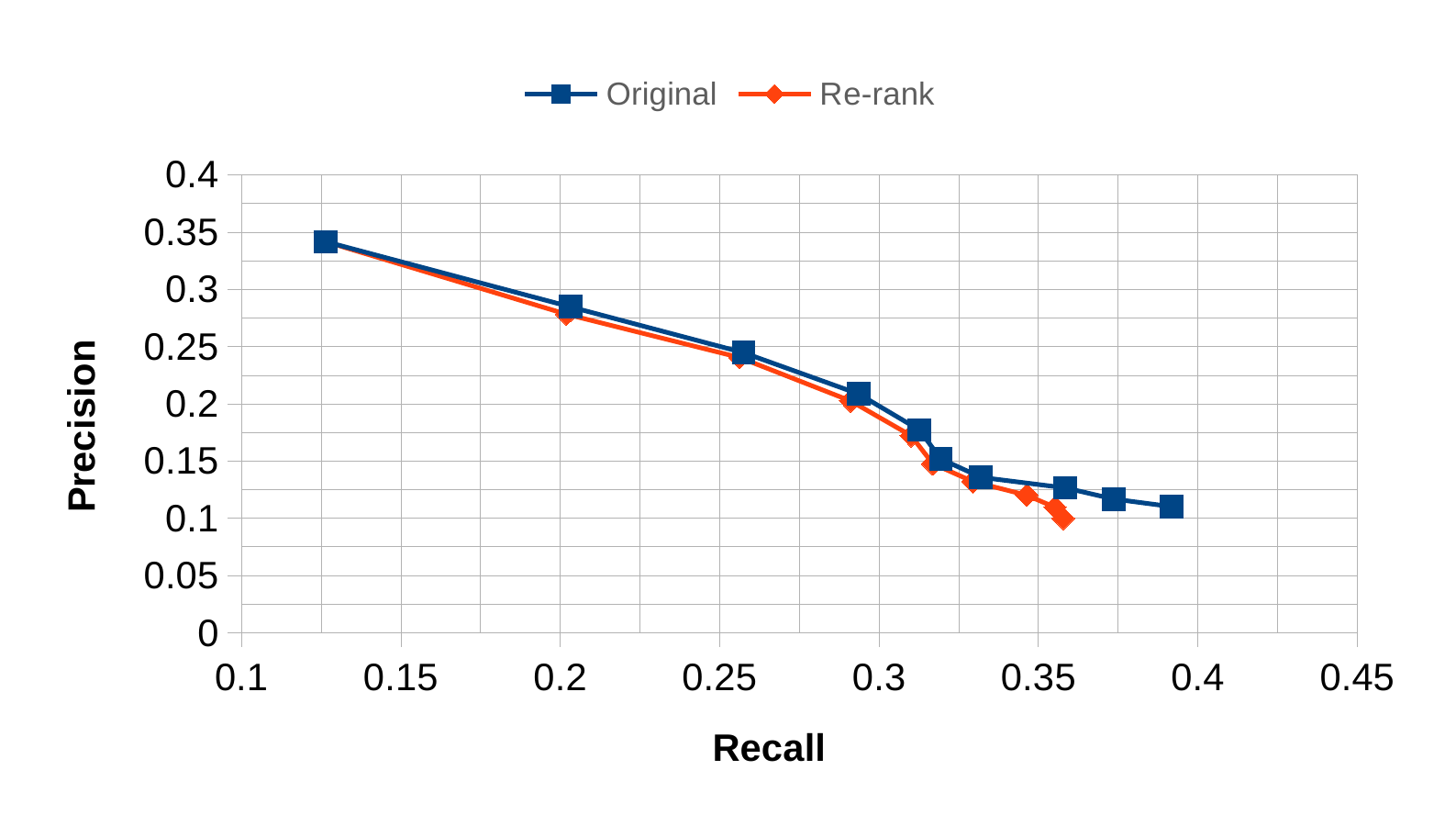}} 
	\end{tabular}
	\caption{Results obtained on the recommendations by running \CR with DS$_1$.}
	\label{fig:CrossRecMetricsDS1}
\end{figure*}

\begin{figure*}[t!]
	\centering    
	\vspace{-.2cm}
	\begin{tabular}{c c c}			
		\subfigure[Expected popularity complement]{\label{fig:EPC_CrossRec_DS2}\includegraphics[width=56mm]{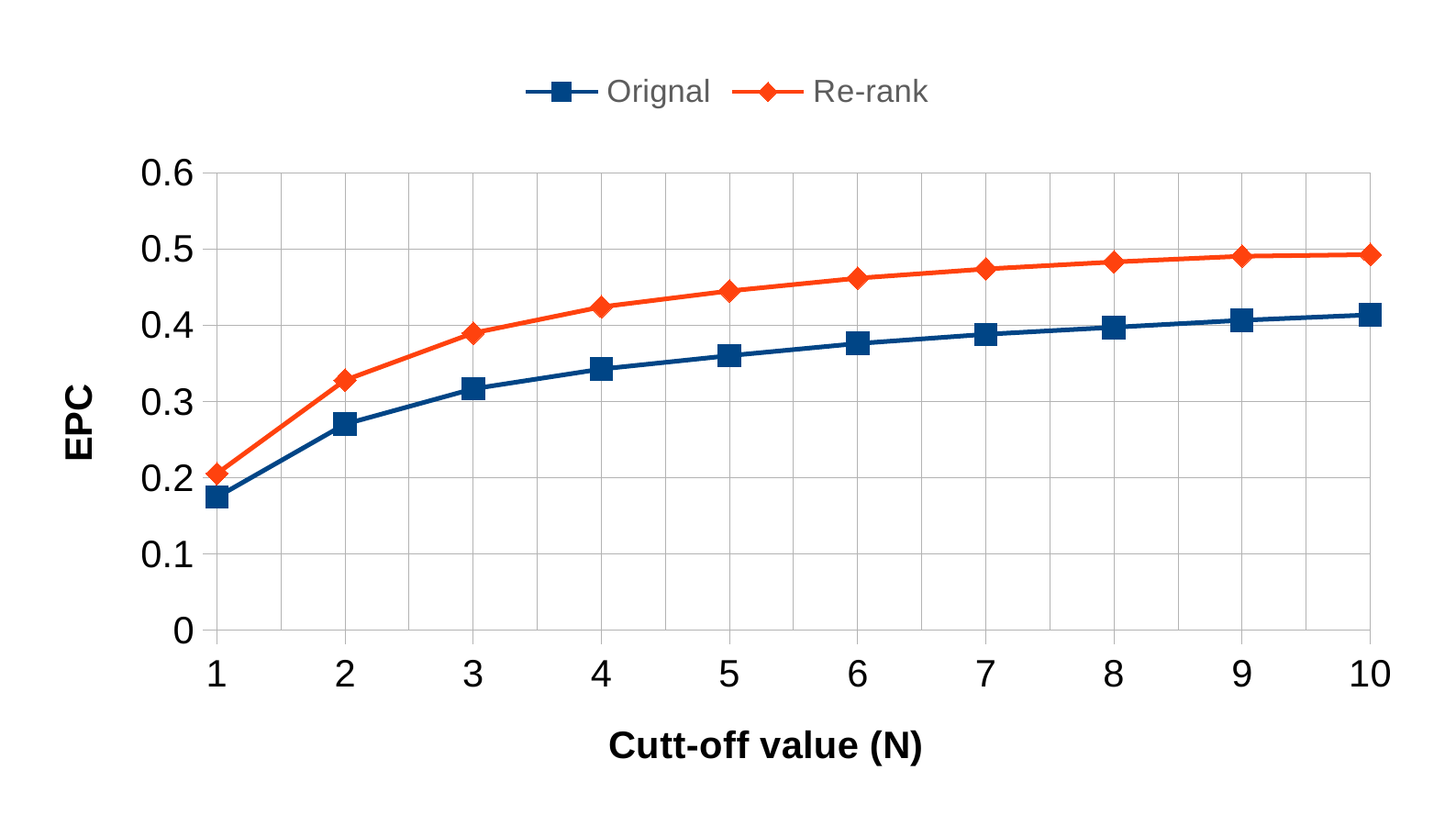}} &		
		\subfigure[Coverage]{\label{fig:Catalog_CrossRec_DS2}\includegraphics[width=56mm]{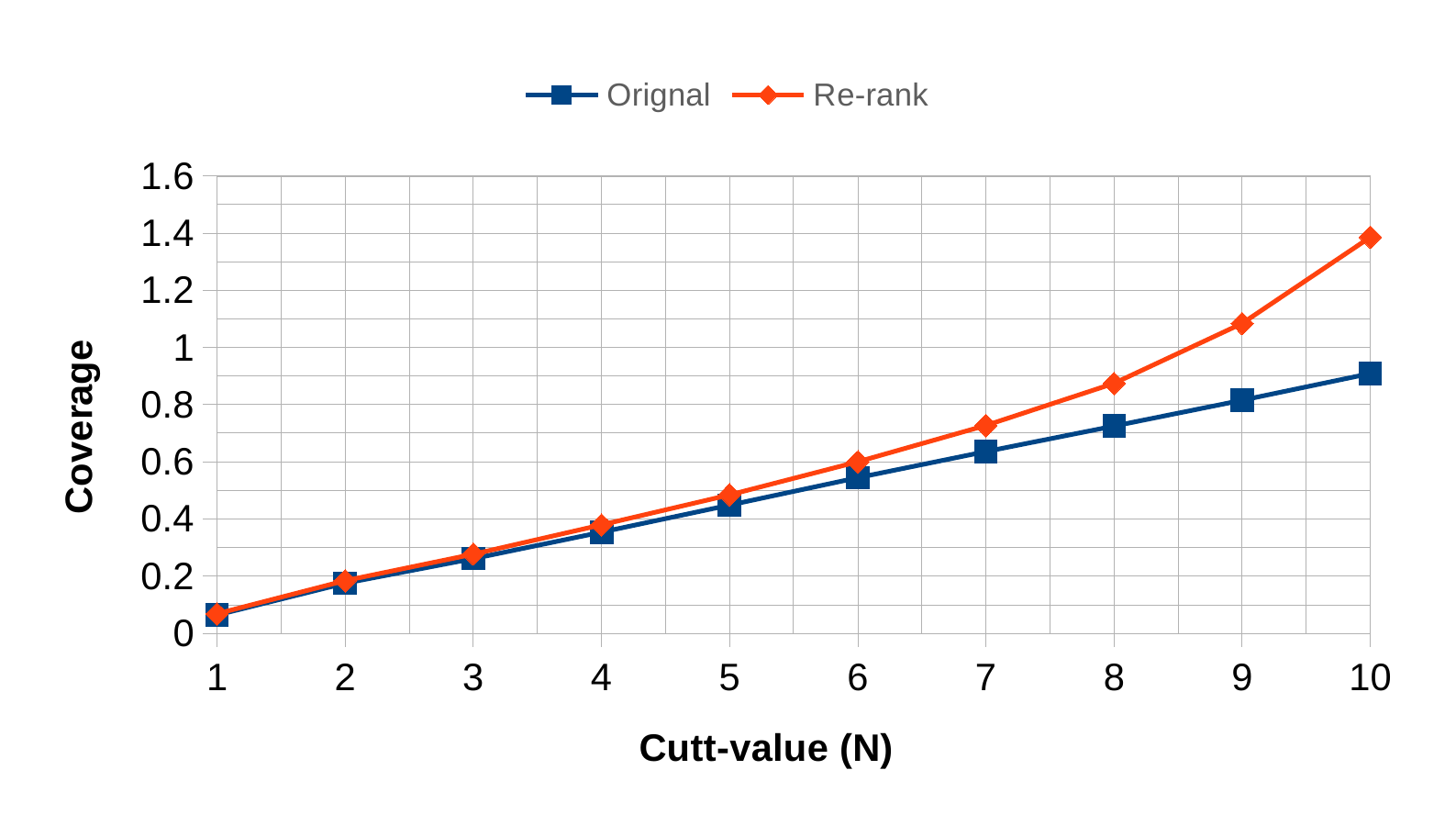}} &
		\subfigure[Precision and recall]{\label{fig:PrecisionRecall_CrossRec_DS2}\includegraphics[width=56mm]{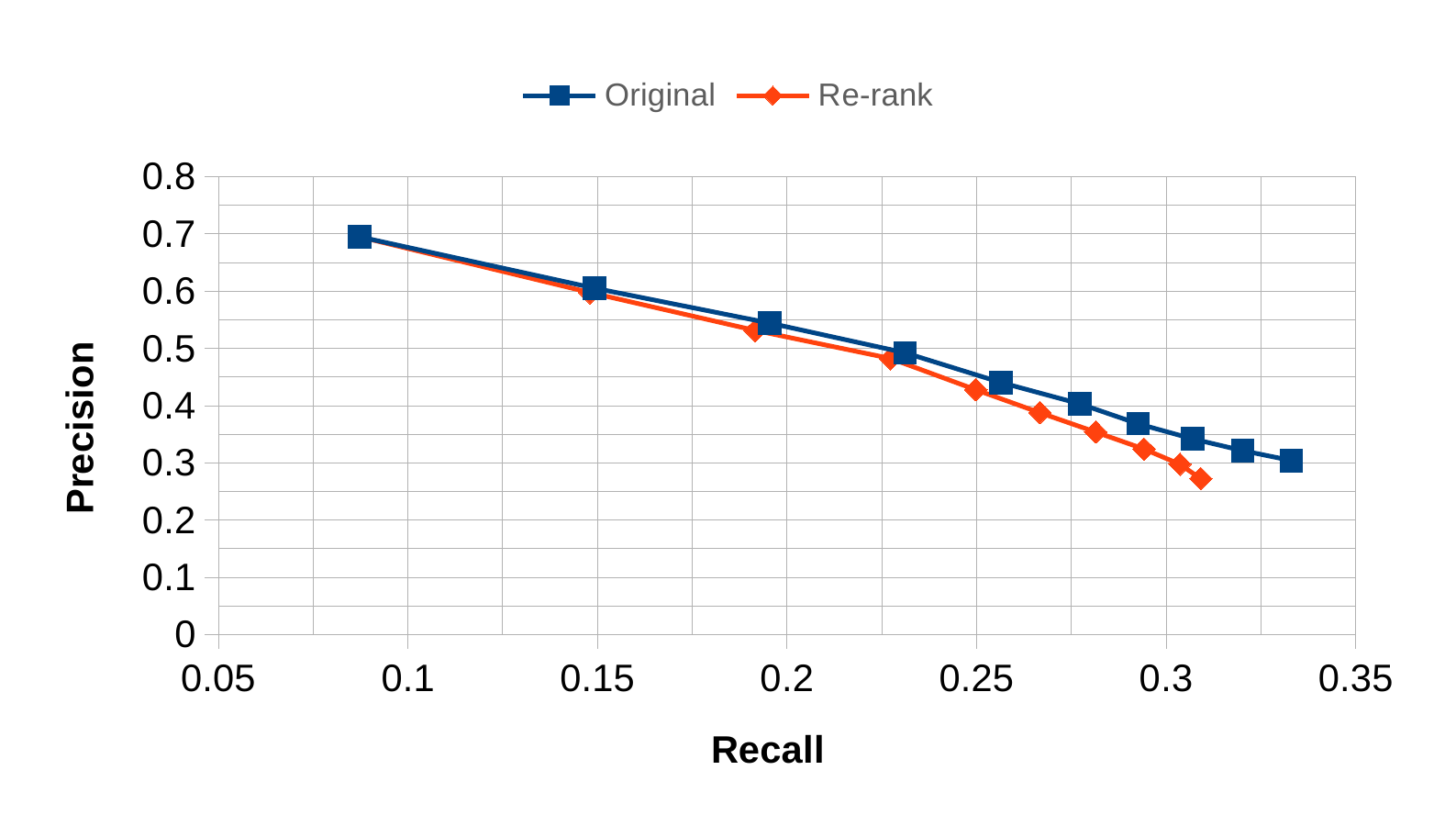}} 
	\end{tabular}
	\caption{Results obtained on the recommendations by running \CR with DS$_2$.}
	\label{fig:CrossRecMetricsDS2}
\end{figure*}




Based on RQ$_2$ results, we found that generally, \LS recommends a wide range of libraries instead of focusing on only popular items. Thus, the recommendations provided by \LS are less prone to popularity bias than those by other systems. Nevertheless, RQ$_2$ also shows that popularity bias is still an issue for TPL RSSEs, triggering the need for proper countermeasures. Due to these reasons, we do not choose \LS for experimenting with the re-ranking technique, which needs a considerably biased recommendation list as input. In turn, \GR returns a short recommendation list, which gives no opportunity for re-ranking by rewarding libraries in the lower part of the ranked list.
Thus, from the \numSys systems in RQ$_2$, we decided to apply re-ranking on \LR and \CR, which usually return a longer ranked list of items. We experimented with the recommendations obtained by the two systems on DS$_1$ and DS$_2$. After a first empirical evaluation, we noticed that on DS$_1$, \LR always recommends very popular items, even late in the list, and it never provides any rare libraries (see Fig.~\ref{fig:LibRecD1_10} and Fig.~\ref{fig:LibRecD1_20}). Therefore, \LR could only benefit from re-ranking on DS$_2$.

The results are depicted in Fig.~\ref{fig:LibRecEvalutionMetrics}, Fig.~\ref{fig:CrossRecMetricsDS1}, and Fig.~\ref{fig:CrossRecMetricsDS2}. Overall, we witness a similar pattern in the impact of the proposed ranking mechanism on the two RSSEs according to the three considered metrics. Thus, we analyze the performance by reporting the results according to each metric as follows.

\vspace{.05cm}
\noindent
$\rhd$~\textbf{EPC.} Following Equation~\ref{eqn:EPC}, a larger EPC means  better novelty, as shown in Fig.~\ref{fig:EPC_LibRec_DS2}, Fig.~\ref{fig:EPC_CrossRec_DS1}, and Fig.~\ref{fig:EPC_CrossRec_DS2} it is evident that by moving rare libraries upper in the ranked list, we are able to improve the EPC scores for the recommendations of both systems, by all the 
cut-off values (N). Especially for \LR, there is a sharper increase in EPC compared to \CR. For instance, as shown in Fig.~\ref{fig:EPC_LibRec_DS2}, when N=10 we obtain an EPC of 3.0 after re-ranking, which is greater than EPC=2.0 the corresponding value achieved on the original list.

\vspace{.05cm}
\noindent
$\rhd$~\textbf{Coverage.} Fig.~\ref{fig:Catalog_LibRec_DS2}, Fig.~\ref{fig:Catalog_CrossRec_DS1}, and Fig.~\ref{fig:Catalog_CrossRec_DS2} show that there is a light improvement in the coverage of the recommendations once the lists have been re-ranked. 
The gain is more evident for the bottom part of the 
list. This is understandable as the proposed technique tends to pad rare items at the end of each list. As a consequence, re-ranking increases coverage, recommending a wider set of libraries, thus improving diversity~\cite{NGUYEN2019110460}.

\vspace{.05cm}
\noindent
$\rhd$~\textbf{Precision and Recall.} 
We investigate 
how the prediction accuracy changes after the re-ranking process by sketching the precision-recall curves following the cut-off values N in Fig.~\ref{fig:PrecisionRecall_LibRec_DS2}, Fig.~\ref{fig:PrecisionRecall_CrossRec_DS1}, Fig.~\ref{fig:PrecisionRecall_CrossRec_DS2}. Essentially, a curve close to the right upper corner represents both higher precision and recall, corresponding to more accurate predictions~\cite{di_rocco_development_2021}. We can notice how the curve representing the performance after re-ranking stays a bit below the one for the original list. This corresponds to a slight setback in precision and recall, once the re-ranking has been conducted. The results indicate that, by promoting the long tail items upper in the list, we encounter false positives in several projects. This is understandable as promoting rare libraries brings true positives to a handful of projects, and false positives to others.



\vspace{.1cm}
\begin{shadedbox}
	\small{\textbf{Answer to RQ$_3$.} On the considered systems, \ie \LR and \CR, the re-ranking technique helps mitigate popularity bias and increase novelty in the recommendation results. Nevertheless, it introduces a setback in prediction accuracy, necessitating further investigation to solve the popularity bias problem properly.} 
\end{shadedbox}

\section{Discussion}
\label{sec:Countermeasures}
This section discusses possible implications and threats to the validity of our findings.

\subsection{Implications} 



RQ$_1$ and RQ$_2$ reveal that software engineering research has not properly dealt with popularity bias in TPL RSSEs, triggering the need for effective counteracting mechanisms. 
Looking at other domains, we see that there are three major methods to combat bias in general, \ie the pre-processing, in-processing, and post-processing paradigms~\cite{doi:10.1089/big.2016.0048}. We assume that they can also be adopted in Software Engineering for the same purpose, as it was also advocated by Chakraborty \etal \cite{ChakrabortyM0M20}, though not directly for RSSEs. 
Also, as it can be seen from Section~\ref{sec:Results}, He \etal~\cite{9043686} conceived \LS--the very first TPL RSSE to mitigate the abundance of highly frequent libraries, following the \emph{in-processing} approach. \LS attempts to neutralize the bias caused by the popularity of TPLs using an adaptive weighting mechanism. However, while it can diversify the recommendations, \LS suffers from a low accuracy (see RQ$_2$ in Section~\ref{sec:RQ2}). This implies that there is still room for improvement, \ie conceiving more effective in-processing techniques for RSSEs. 

\revised{Through RQ$_2$, we can notice that the characteristics of datasets are a contributing factor in the ability of \LS to mitigate popularity bias. In fact, DS$_3$ contains a considerably large number of projects (56,091), but only a small number of libraries (762). In contrast, compared to DS$_3$, DS$_2$ has a lower number of projects (5,200), but its number of libraries is substantially higher (31,817). In this respect, the distribution of the libraries across the projects in DS$_3$ is much denser than that of DS$_2$, as the former features projects from the same ecosystem (Android). It is then necessary to perform an in-depth analysis of how the characteristics of a dataset impact on the ability of TPL recommender systems to deal with popularity bias.} 

In this work, we derived a \emph{post-processing} technique following an algorithm for  diversifying Web search results, defusing popularity bias without touching the internal design of the considered systems. On the one hand, the technique helps improve diversity in the recommendations of two systems. On the other hand, the improvement is rather modest, and there is still a slight decrease in accuracy. \revised{Our findings suggest that 
further research should be conducted to propose effective countermeasures. We suppose that it is crucial to consider 
additional factors, \eg the degree of specificity (to certain solutions) of a TPL, when it comes to library recommendation.} 


\emph{Pre-processing} techniques can also be useful for TPL RSSEs, \eg by considering factors related to peculiar, solution-specific aspects of a project that, as pointed out in Section \ref{sec:Background} are neglected, and that would help reward certain libraries. From RQ$_1$, we see that no approach specifically employs pre-processing to cope with popularity bias in RSSEs.


\subsection{Threats to validity} \label{sec:Threats}


Threats to \emph{construct validity} are related to the relationship between theory and observation. This concerns the setting 
used to query the RSSEs. We simulated a basic scheme where given a testing project, half of its libraries are used as a query, and the remaining is ground-truth data. Such a scenario resembles a practical use, in which developers search for suitable TPLs once they have invoked some libraries in their project.


Threats to {\em internal validity} are the confounding factors internal to our study that might have an impact on the results. 
For the literature analysis, given a paper that needs to be reviewed, we only classified it by its title and abstract. A possible threat is that popularity bias can be addressed elsewhere in the paper, \eg approach/evaluation. In this case, we may miss relevant studies. Another threat is related to the considered venues, 
\ie we might only cover some of the relevant conferences and journals. As shown in Section~\ref{sec:Collection}, we considered major software engineering venues where state-of-the-art research in RSSEs is presented. 
We used their original implementations to experiment on the \numSys selected RSSEs. 

Threats to \emph{external validity} concern the generalizability of our results. The conclusion drawn from the experiments is valid for the \numSys considered systems, and it may no longer apply to other TPL RSSEs. We anticipate that expanding the evaluation on more systems will help us further validate our hypothesis. 
To aim for diversity in the training data, we used datasets that cover two major domains, \ie generic software and Android applications, which exhibit diverse characteristics to simulate real-world data.



\section{Related work}
\label{sec:RelatedWork}
The literature analysis in Section~\ref{sec:RQ1} shows that popularity bias has not received adequate attention from the 
community. In this section, we further review related work concerning fairness in Software Engineering and tackling bias in recommender systems in other domains.

\subsection{Fairness in Software Engineering}
Brun and Meliou~\cite{10.1145/3236024.3264838} investigated the potential impact of biased data in the critical software engineering phases, \ie requirement specification, system design, testing, and verification.
Their investigation indicated that a comprehensive classification of software biases remains an open challenge. Similarly, Verma and Rubin~\cite{verma_fairness_2018} studied fairness in the context of algorithmic classification. Their evaluation with a state-of-the-art dataset and logistic regression classifiers shows that fostering software fairness is challenging since the tested classifier is defined as fair according to the chosen definition. 
Spoletini \etal \cite{spoletini_bias-aware_2018}  proposed an initial set of bias-aware guidelines, providing practical instructions when it comes to dealing with bias in software engineering algorithms. 

Chakraborty \etal \cite{ChakrabortyM0M20} advocated for the need for fairness analysis and testing in machine learning software. They propose an approach named Fairway that removes bias during pre-processing (\ie before training) and in-processing (\ie during training), and uses multi-objective optimization to avoid that fairness compromises the machine learning performance. In  subsequent work, Chakraborty \etal \cite{ChakrabortyMM21} 
proposed a fairness-aware data rebalancing approach, named Fair-SMOTE,  that leverages situation testing to balance fair-sensitive labels, outperforming previously-proposed approaches including the previously-proposed Fairway. %
While Chakraborty \etal deal with fairness by handling fair-related attributes, in RQ$_3$ of this work we leverage a technique inspired  by Web search~\cite{10.1145/1772690.1772780} to diversify TPL recommendations.
Last but not least, multi-objective approaches like the one of Chakraborty \etal can also be applied in mitigating popularity bias in TPL RSSEs, however, the goal is different (\ie reranking rare items that could be useful for some projects instead of coping with fairness-related features).

\subsection{Recommender system bias in other domains}

Kowald and Lacic~\cite{kowald_popularity_2022} investigated popularity bias in collaborative filtering (CF) multimedia recommender systems. They first selected four datasets composed of users and corresponding media items and then 
 evaluated four CF algorithms in terms of popularity bias.
 Their results showed that users with a low preference to popular items receive significantly worse recommendations compared to the others, meaning that all the considered systems are prone to popularity bias.

The CPFair framework~\cite{naghiaei_cpfair_2022} leverages an optimization-based approach to support the multi-stakeholders fairness requirements in the multimedia domain. Given a list of items provided by an unfair recommender, CPFair exploits a re-ranking post-hoc algorithm to integrate the fairness requirements and remove the bias in the initial set of recommended items.


d'Aloisio \etal~\cite{DALOISIO2023103226} proposed a data-agnostic Debiaser for Multiple Variables (DEMV) to mitigate biases in multi-class classification problems. 
The approach alleviates bias using random sampling to remove or duplicate elements until the observed size converges to the expected one. %
FairSR~\cite{li_fairsr_2022} is a fairness-aware sequential recommender system that exploits knowledge graphs to increase interaction fairness, \ie users belonging to different groups interact with the items equally. 

Li \etal~\cite{li_fairgan_2022} proposed a recommender system based on Generative Adversarial Networks, called FairGAN, to handle fairness with implicit user feedback. The tool extracts feedback from user interactions and ranks the items accordingly. Afterward, it generates fairness signals to enforce the exposure of items by satisfying different fairness criteria.   

While previous work has studied recommender bias in other domains, ours is the first thorough investigation of popularity bias for RSSEs, and specifically for TPL RSSEs, reviewing the existing literature, by comparing the behavior of four TPL RSSEs, and attempting to mitigate their popularity bias.

\section{Conclusion and future work}
\label{sec:Conclusion}

In this paper, we performed both a qualitative and quantitative evaluation to study the presence of popularity bias in TPL RSSEs. A literature review on major SE venues reveals that the issue of dealing with popularity bias has not received enough attention from the community. The finding is further confirmed with an empirical evaluation on \numSys TPL RSSEs, \ie three among the considered systems recommend highly frequent libraries to projects. One system, while being able to undermine the effect of recommending popular libraries, suffers from a low prediction accuracy. Altogether, we see that state-of-the-art research in software engineering overlooks the problem of popularity bias in third-party library recommender systems, leaving a research gap that needs to be properly filled.

For future work, we plan to conceive more effective mechanisms for counteracting popularity bias, increasing fairness in the recommendations, \eg by means of a multi-objective approach, while keeping high accuracy. 
More importantly, we assume that it is necessary to study the issue of popularity bias in other RSSEs, for instance, API and code recommender systems as well as pre-trained models for code search. 

\section*{Acknowledgments}
This work has been partially supported by the EMELIOT national research project, which has been funded by the MUR under the PRIN 2020 program (Contract 2020W3A5FY).

\balance
\bibliographystyle{IEEEtranS}
\bibliography{IEEEabrv,main}

\end{document}